\documentclass{aa}  

\usepackage{graphicx}
\usepackage{txfonts}
\usepackage{hyperref}
\usepackage[dvipsnames]{xcolor}

\newcommand{\Msun}{\,M$_\odot$}
\newcommand{\Rsun}{\,R$_\odot$}

\newcommand{\tento}[1]{$10^{#1}$}
\newcommand{\timestento}[2]{$#1 \times 10^{#2}$}
\newcommand{\Msunyr}{\,$\mathrm{M}_{\odot}$\,$\mathrm{yr}^{-1}$}

\newcommand{\revision}[1]{#1}

\newcommand{\action}[1]{}
\newcommand{\rev}[1]{#1}
\newcommand{\LEt}[1]{}

\begin{document}

\title{A systematic study of super-Eddington \revision{layers in the envelopes of massive stars}}

\subtitle{}

\author{Poojan Agrawal\inst{1,2,3} \thanks{E-mail: pagrawal@astro.swin.edu.au}
\and
Simon Stevenson\inst{1,2}
\and
Dorottya Sz\'ecsi\inst{4}
\and
Jarrod Hurley\inst{1,2}
          }

\institute{Centre for Astrophysics and Supercomputing, Swinburne University of Technology, Hawthorn, VIC 3122, Australia
\and 
OzGrav: The ARC Centre of Excellence for Gravitational Wave Discovery, Hawthorn, VIC 3122, Australia
\and 
McWilliams Center for Cosmology, Department of Physics, Carnegie Mellon University, Pittsburgh, PA 15213, USA
\and 
Institute of Astronomy, Faculty of Physics, Astronomy and Informatics, Nicolaus Copernicus University, Grudziądzka 5, 87-100 Toruń, Poland}

   \date{Received September 15, 1996; accepted March 16, 1997}

 
  \abstract
   {The proximity to the Eddington luminosity has been attributed as the cause of several observed effects in massive stars. 
    Computationally, if the luminosity carried through radiation exceeds the local Eddington luminosity in the low-density envelopes of massive stars, it can result in numerical difficulties, inhibiting further computation of stellar models. This problem is exacerbated by the fact that very few massive stars are observed beyond the Humphreys-Davidson limit, the same region in the Hertzsprung–Russell diagram where the aforementioned numerical issues relating to the Eddington luminosity occur in stellar models. 
}
   {One-dimensional\LEt{ Avoid beginning a sentence with an abbreviation, and do not begin with a number (unless written out), a formula, or symbol.} stellar evolution codes have to use pragmatic solutions to evolve massive stars through this computationally difficult phase. In this work, we quantify the impact of these solutions on the evolutionary properties of massive stars. }
   {We used\LEt{ A\&A uses the past tense to describe specific methods used in a paper, and the present tense to describe general methods and the findings of recent papers. See Sect. 6 of the language guide https://www.aanda.org/for-authors/language-editing/6-verb-tenses.} the stellar evolution code MESA with commonly used input parameters for massive stellar models to compute the evolution of stars in the initial mass range of 10--110\Msun{} at one-tenth of solar metallicity.}
   {We find that numerical difficulties in stellar models with initial masses greater than or equal to 30\Msun{} cause these models to fail before the end of core helium burning. 
    Recomputing these models using the same physical inputs but three different pragmatic solutions to treat the numerical instability, we find that the maximum radial expansion achieved by stars can vary by up to 2000\Rsun{}, while the remnant mass of the stars can vary by up to 14\Msun{} between the sets. These differences can have implications on studies such as binary population synthesis.
 }
   {}

   \keywords{Stars: massive --  supergiants -- Stars: evolution -- Stars: black holes -- gravitational waves -- methods:numerical}

   \maketitle
%

\section{Introduction}
Stars more massive than about 9\Msun{} are key to several astrophysical processes. During their lives, they enrich their surroundings with ionizing flux and nuclear-processed material while altering the dynamics of their host systems. At the end of their lives, these massive stars again expel a copious amount of radiation and metal-rich matter in the form of supernovae, leaving behind compact remnants: neutron stars and black holes.
Furthermore, mergers of these compact remnants result in gravitational wave \citep[][]{Abbott:2016,Abbott:2017NSmerger} emission and can also lead to the formation of rare elements \citep[][]{Kasen2017}.
Therefore, a better understanding of how these stars evolve is crucial in comprehending their contribution to the evolution of star clusters and galaxies.  

The evolution of massive stars is typically modeled using one-dimensional (1D) stellar evolution codes. 
In the last few decades, these stellar evolution codes have progressed a lot and so has our understanding of massive stars. Together with the advances in our observing capabilities \citep[][]{Evans:2011,Simon-Diaz2015IACOB,Wade2014MiMes,Abbott:2016}, the development of sophisticated numerical methods for simulating physical processes and newer input data in the form of opacity tables and nuclear reaction rates has led to the development of modern and improved stellar structure and evolution codes \citep[][]{Langer:2012, Ekstroem2020}.

Even with these new capabilities, 1D modeling of massive stars is limited by a number of approximations. Several evolutionary properties of massive stars such as mass-loss rates \citep{Smith:2014,Renzo:2017}, nuclear reaction rates \citep{Heger:2002,Fields:2018},\LEt{ A\&A uses the serial (Oxford) comma between three or more items in a list to avoid confusion. Also use commas after introductory sentences of three or more words. Commas are not necessary between just two parallel items in a sentence.} and rotation \citep[][]{Heger:2000a,Maeder2009} remain uncertain.
The high mass of these stars makes it feasible for \LEt{ or "for sophisticated physical processes".}\action{changed "to sophisticated physical processes"} sophisticated physical \rev{processes} to operate in these stars, but their short lives makes it difficult to obtain observational constraints. 

One such process is the treatment of convective transport of energy through the\LEt{ Names and the full version of acronyms do not always require each word to be capitalized. Please refer to Sect 2.2 of the language guide and amend caps throughout.} mixing length theory \citep[MLT; ][]{BohmVitense1958}. 
In this theory, energy is transported through fluid elements supported by buoyancy forces. These elements travel over a radial distance known as the mixing length, after which they dissolve in their surroundings.  
MLT assumes hydrostatic equilibrium in stars and only depends on local conditions (i.e., local values of pressure, density, etc.), without taking other parts of the star into account. Time-dependency and nonlocal treatments are included through ad \LEt{ We follow the CMOS rule of not hyphenating modifiers composed of words from foreign languages.}hoc methods such as convective overshoot, semiconvection, and diffusion \citep[e.g.,][]{Renzini1987,Kippenhahn2012}.

The simplicity of MLT makes it a popular choice for many stellar evolution codes. 
While MLT gives fairly good results in the deep interiors of stars where density is high and convection is nearly adiabatic (with negligible radiative losses), its limitations start becoming apparent in low-density environments where the convection is highly superadiabatic and prone to radiative losses \citep{Maeder2009}. 
For example, in the low-density envelopes of massive stars, convection, as given by MLT, is inefficient. Furthermore, changes in the elemental opacity as the star evolves can cause the radiative luminosity to exceed the local Eddington luminosity and can lead to the formation of density and gas pressure inversions in the subsurface layers.

The proximity to the Eddington luminosity and the associated density inversions have been attributed as the source of several instabilities in massive stars, for example, the dynamical instability \citep{Stothers:1993}, the convective instability \citep{Langer1997}, and the strange-mode instability \citep{Saio1998, Saio2013}.
Observationally, these have been linked to stellar variability phenomena such as stochastic low-frequency photometric variability \citep{Pedersen2019, Bowman2020}, spectroscopic macroturbulence \citep{simon-diaz2014,Simon-Diaz2017}, and episodic mass ejection behavior in luminous blue variables \citep[LBVs:][]{Bestenlehner2014, Grafener:2021}.

From a numerical perspective, the presence of density inversions in the inflated envelopes of supergiant stars requires 1D stellar evolution codes to take prohibitively short time steps (on the order of hours and minutes), leading to convergence issues \citep{Maeder1987,Alongi1993,Paxton2013}.
Evolving stars past these numerical difficulties in the supergiant phase has been a long-standing challenge for the 1D stellar evolution approach. As shown in \citet{Agrawal:2022} (hereafter Paper I), stellar evolution codes often resort to numerous pragmatic solutions to evolve stars, 
such as enhancing the convective efficiency \citep[e.g.,][]{Ekstroem:2012} or limiting temperature gradients such that the density gradient is always positive \citep[e.g.,][]{Cheng:2015}.
While these solutions help evolve stars through numerically difficult phases of evolution, they can also modify their surface behavior, such as radius evolution and mass-loss rates. 
 
The different solutions used by 1D codes and their interplay with other physical parameters of massive stars can alter the dynamics of stellar evolution, thereby adding a potential bias to any study aiming to determine the properties of these input parameters from the evolution of a star. While other physical uncertainties in the evolution of massive stars have received considerable attention in a number of studies, the impact of numerical issues has not been explored to the same extent. 

In Paper I, the comparison of stellar models from five different codes revealed large differences in the evolutionary behavior of stars more massive than 40\Msun{} around solar metallicity (at $Z = 0.014$). The maximum radial expansion predicted by different stellar models varied by more than 1000\Rsun{} and the predictions of remnant mass varied by 20\Msun{}.
However, the stellar models also had different physical inputs besides the pragmatic treatment of density inversions arising due to the Eddington luminosity, making it difficult to untangle the impact of this process from other inputs. There is therefore a need for a systematic study of the impact of these solutions within a single code and single set of assumptions.

In this work, we perform a study of the impact of density inversions on the evolution of massive stars up to 110\Msun{} using consistent input parameters. Since Paper I focused on solar metallicity stars only, we chose a metallicity ten times lower than solar here to demonstrate the impact of density inversions at a metallicity relevant to the progenitors of current gravitational wave observations \citep[e.g.,][]{Stevenson:2017}. 
As we show here, the different pragmatic solutions used by 1D codes can have a non-negligible impact on the evolutionary properties of massive stars. These differences are important, as they can help us explain the formation of gravitational wave progenitors and other observations of stellar populations.

The paper is organized as follows. We provide an overview of the physics related to density inversions in Section~\ref{sec:physics_density_inv}. In Section~\ref{sec:standard_model}, we describe our standard or default set of models with MESA and discuss the effect of density inversions on their completeness. We recompute the models that fail to reach the end of evolution, using three different pragmatic solutions in Section~\ref{sec:model_variations} and compare the impact of these solutions in predicting the different evolutionary properties of massive stars in Section~\ref{sec:implications}. In Section~\ref{sec:observations_hd_limit}, we compare the stellar models from Section~\ref{sec:model_variations} with the observations of massive stars and conclude our study in Section~\ref{sec:conclusion}.

\section{Physics of density inversions in massive stars}     
\label{sec:physics_density_inv}

\LEt{ Single-sentence paragraphs are not allowed.}\action{corrected} In this section, we describe the conditions for the formation of density and gas pressure inversions in stellar envelopes and their impact on modeling the evolution of massive stars.
For a spherically symmetric star containing mass $m(r)$ inside radius $r$ and with radiative opacity $\kappa(r)$ and density $\rho(r)$, the luminosity that can be carried by radiative transport of energy is given by

\begin{equation}
    L_{\rm {rad}}(r)=-\frac{4 \pi r^2 c }{\rho(r) \kappa(r)} \frac{d P_{\rm rad}}{d r}\, ,
    \label{eq:lrad}
\end{equation}
where $P_{\rm rad}$ denotes the radiation pressure and $c$ is the speed of light. 

The Eddington luminosity gives the maximum value of luminosity that can be transported by radiation while maintaining hydrostatic equilibrium \citep{Eddington1926}. 
The expression for the Eddington luminosity is given by
\begin{equation}
    L_{\rm {Edd}}(r)=\frac{4 \pi c G m(r)}{\kappa(r)} \, ,
    \label{eq:ledd}
\end{equation}
where $G$ represents the gravitational constant. 
The total pressure $P$ in the star is the sum of the radiation pressure, $P_{\rm rad}$ and the gas pressure $P_{\rm gas}$.
Using equations\,\ref{eq:lrad} and \ref{eq:ledd} and the equation of hydrostatic equilibrium $d P/d r = -G m(r) \rho(r)/ r^2$, the ratio of radiative luminosity to the Eddington luminosity (Eddington factor) can be defined as

\begin{equation}
    \frac{L_{\rm rad}}{L_{\rm {Edd}}} = \frac{d P_{\rm rad}}{d P} = \left[1+\frac{d P_{\rm gas}}{d P_{\rm rad}}\right] \, .
    \label{eq:gamma}
\end{equation}

Normally, the luminosity transported by radiation $(L_{\rm rad})$ is less than the Eddington luminosity $(L_{\rm Edd})$ inside a star,  and the density and gas pressure of stellar material decrease with the stellar radius ($d \rho/d r<0, d P_{\rm gas}/d r<0)$.
However, during the evolution of the star, changes in the ionization states of various elements can lead to an increase in the opacity $\kappa(r)$. In the low density ($\rho \ll {\rm 1\,g\,cm^{-3}}$) radiation pressure dominated ($P_{\rm rad}/ P \approx 1$) envelopes of massive stars,
an increase in opacity can reduce the local Eddington luminosity below the radiative luminosity $(\frac{L_{\rm rad}}{L_{\rm Edd}}>1)$. 
Since $d P_{\rm rad}/d r$ is always less than 0, when 
$\frac{L_{\rm rad}}{L_{\rm Edd}}>1$, Equation\,\ref{eq:gamma} implies $d P_{\rm gas}/d r>0$, that is\LEt{ e.g./i.e. should be written out in full when part of the main text (not inside parentheses or figure legends). i.e. should be replaced by "that is" or similar when in main text. See the entries for e.g. and i.e. in Section 2.1, "Note 2" of the language guide https://www.aanda.org/for-authors/language-editing/2-main-guidelines.}, a gas pressure inversion. 

From the ideal gas equation of state, $P_{\rm gas}$ can be expressed as a function of $\rho$ and $P_{\rm rad}$.
Using
Equation\,\ref{eq:gamma}, the condition for density inversion ($d \rho/d r>0)$ can therefore be written as 
\begin{equation}
\frac{L_{\mathrm{rad}}}{L_{\mathrm{Edd}}} > \left[1+\left(\frac{\partial P_{\mathrm{gas}}}{\partial P_{\mathrm{rad}}}\right)_{\rho}\right]^{-1}\, 
\label{eq:density_inv}
\end{equation}
\citep{Joss1973, Paxton2013}. 

\revision{Since $\left(\frac{\partial P_{\mathrm{gas}}}{\partial P_{\mathrm{rad}}}\right)_{\rho}$ 
is a always positive, $\frac{L_{\rm rad}}{L_{\rm Edd}}>1$ implies a density inversion in stars.}

The effect of density inversions on the evolution of massive stars is complex and remains an active field of research \citep{Mihalas:1969,Ergma1971,Langer1997,Owocki2004, Cantiello2009, Sanyal:2015}. 
\citet{Grafener:2012} found that an increase in the radiative luminosity near the opacity peak due to the iron-group elements (at $\sim$\timestento{2}{5} K) leads to the formation of an "inflated envelope"\LEt{ US convention uses double quotation marks to indicate a special use of a word or phrase.} containing density inversions \citep[also see][]{Ishii:1999, Petrovic:2006,Kohler:2015}. 
In this state, the star has an extended radiative envelope with a relatively small convective core. As pointed out by \cite{Sanyal:2015}, these inflated stars are different from the classical red supergiants, as envelope inflation does not require hydrogen shell burning and can even occur while the star is on the main sequence.

\revision{For hydrogen-rich stars, the formation of the inflated envelope reduces the opacity and therefore the Eddington factor, ${L_{\rm rad}}/{L_{\rm Edd}}$, stays less than or close to the unity.
However, as the envelope expands, its outer layers become sufficiently cool to encounter even steeper opacity bumps due to helium ionization (at $\sim$\timestento{2}{4} K) and hydrogen recombination (at $\sim$\tento{4} K). Further expansion no longer reduces ${L_{\rm rad}}/{L_{\rm Edd}}$, and a density inversion forms in the outer layers of the star.} 

\revision{The presence of these inversions in the convectively-inefficient and radiation-dominated envelope of massive stars ($P_{\rm gas}/ P \ll 1$) leads to numerical instabilities \citep{Maeder1987,Maeder:1992} 
and an additional mode of energy transport (such as turbulent convection) may be needed to carry the energy \citep{Jiang2015,Schultz2020}. 
Without proper treatment of these instabilities, 1D stellar evolution codes find it difficult to obtain the solution of the stellar structure equations. They are, therefore, forced to adopt exceedingly small time steps and struggle to complete the evolution of these stars \citep{Paxton2013}.}



\begin{figure*}
    \centering
    
    \begin{tabular}{cc}
    
     \includegraphics[width=0.485\textwidth]{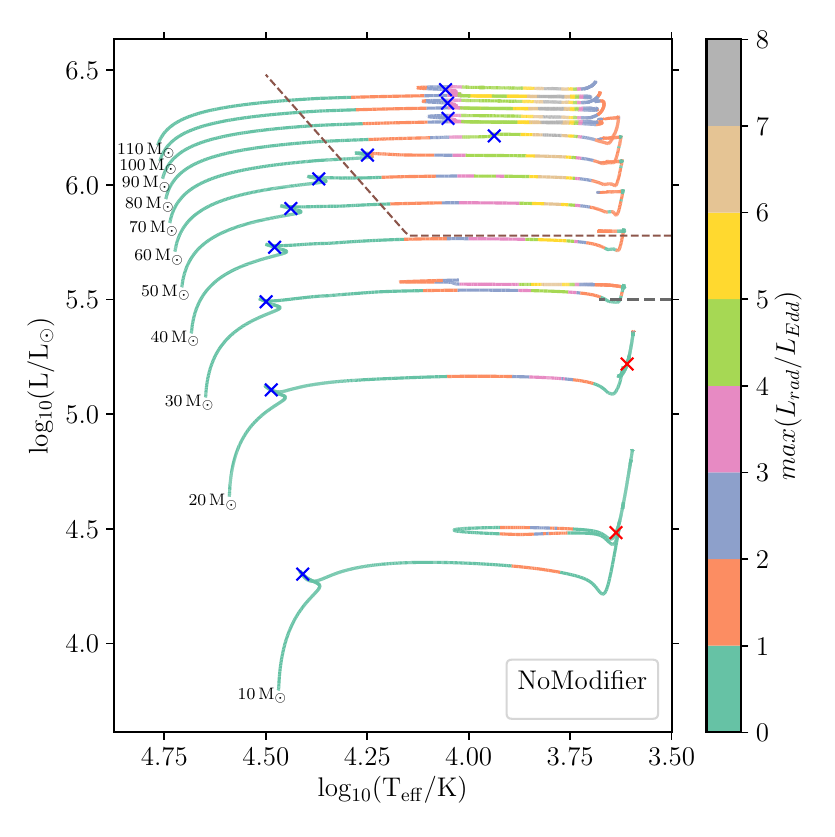}
     &
     \includegraphics[width=0.5\textwidth]{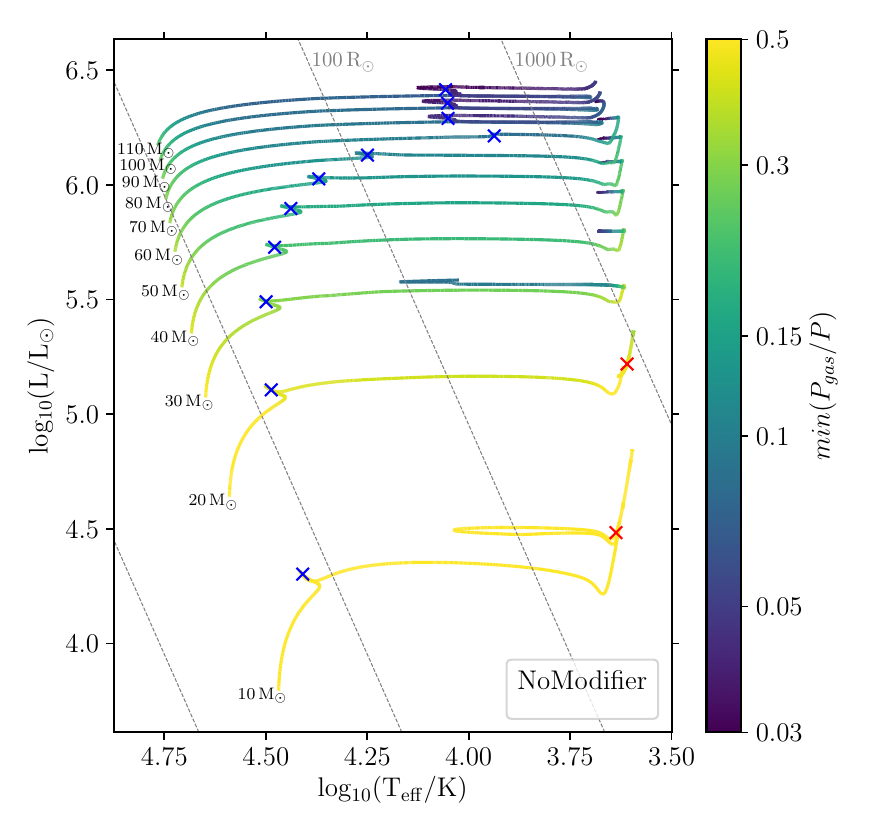}
    
    \\
    \end{tabular} 
    \caption{Hertzsprung-Russell (HR) diagram showing stellar models from the standard set colored by the maximum of ${L_{\rm rad}}/{L_{\rm Edd}}$ (left panel) and the minimum of ${P_{\rm gas}}/{P_{\rm total}}$ (right panel).
    The blue cross marks the end of core hydrogen burning and the red cross marks the end of core helium burning (where applicable). The brown dashed line in the left panel denotes the position of the observational \citet{Humphreys:1979} limit beyond which few stars are observed, while the gray dashed line signifies the luminosities of the brightest red supergiants as inferred by \citet{Davies:2018}. Large values of ${L_{\rm rad}}/{L_{\rm Edd}}$ and small values of ${P_{\rm gas}}/{P_{\rm total}}$  in the envelopes of stars with initial masses 30\Msun{} and above causes the evolution of these stars to become halted at ${\rm log T_{eff}/K \approx 3.7}$, and their models fail to even finish core helium burning.}
  
    \label{fig:Hr_nomodifier}
\end{figure*}

\begin{figure}
    \centering
    \includegraphics[width=\columnwidth]{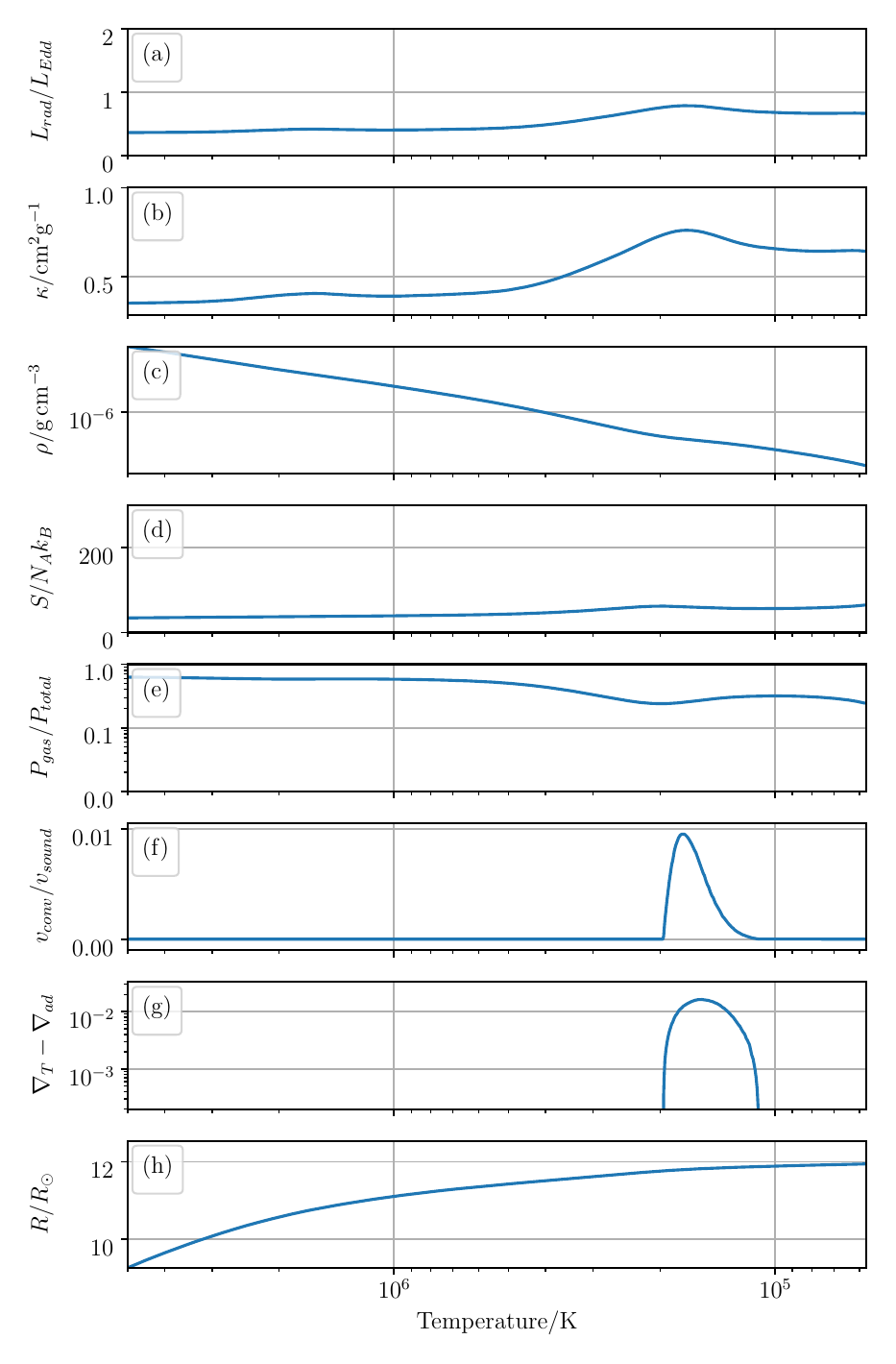}
    \caption{Temperature profile of a 110\Msun{} star at the ZAMS evolved using the standard set. Opacity peaks due to partial ionization states of iron are present at \timestento{1.5}{6} and \timestento{1.8}{5}\,K. However, \revision{${L_{\rm rad}}/{L_{\rm Edd}}$} is less than 1 throughout the star and the evolution of the star proceeds smoothly. \revision{For clarity, only the outermost layers of the star (with $T \leq\,$\timestento{5}{6}\,K) containing the various opacity peaks have been plotted.} See Section~\ref{sec:results} for details of each panel.}
    \label{fig:profile_nomodifier_513}
\end{figure}

\begin{figure}
    \centering
    \includegraphics[width=\columnwidth]{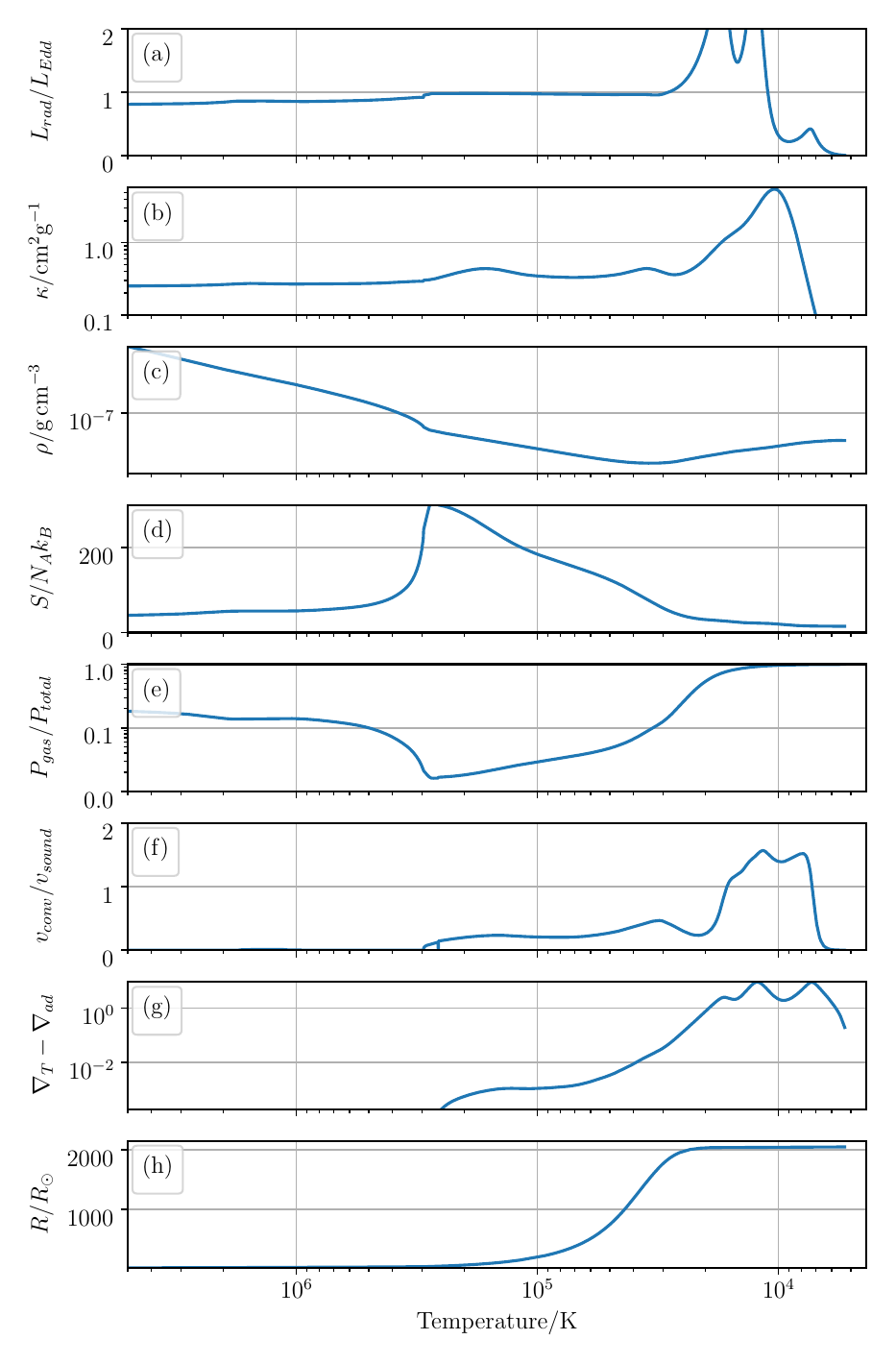}
    \caption{Temperature profile of \revision{the outermost layers of a} 110\Msun{} star at the end of the simulation, evolved using the standard set. 
    The high luminosity of the star combined with the peak in opacity due to hydrogen and helium ionization around \tento{4}\,K causes \revision{${L_{\rm rad}}/{L_{\rm Edd}}$} to exceed unity. This causes density and gas pressure inversions at \timestento{3}{4}\,K. However, the low density of the environment renders convection inefficient and the star struggles to evolve despite reaching supersonic convective velocities.}
    \label{fig:profile_nomodifier_4311}
\end{figure}

\section{Stellar models: The standard set}
\label{sec:standard_model}

\LEt{ Only the first word of titles and subheadings should be capitalized (excluding proper nouns).}Module for Experiments in Stellar Astrophysics
\citep[MESA:][]{Paxton2011,Paxton2013,Paxton2015,Paxton2018,Paxton2019} is an advanced one dimensional stellar structure and evolution code.
MESA solves the coupled differential equations of the stellar structure simultaneously with the energy transport equation for all mass cells from the surface to the center. In this work, we make use of version 11701 of MESA. 

\subsection{Physical ingredients}
\label{sec:physical_ingredients} 

Convection is treated as a diffusive process with convective boundaries determined by the Ledoux criterion. To account for convection due to varying opacity in layers, we use the \citet{Henyey1965} treatment of MLT with mixing length parameter, $\alpha_{\rm MLT}$ = 1.82.
Semiconvection is implemented using \citet{Langer1985} with $\alpha_{\rm semiconvection} =1.0$ while a small amount of Thermohaline mixing is used ($\alpha_{\rm th} = 2.0$) following \citet{Kippenhahn1980}.
For convective overshooting we follow the calibration adopted by \citet{Brott:2011}: step overshoot with f = 0.335 and f0 = 0.05. We only apply overshooting in the core and not in the convective envelope. 

Mass loss is modeled using the commonly used MESA "Dutch" scheme \citep{Glebbeek:2009} with the mass loss scaling factor, ${\rm \eta_{Dutch}=1.0}$. The scheme follows \citet{Vink:2000,Vink:2001} for hot winds, \citet{deJagerandNieuwenhuijzen:1988} for cool winds and \citet{NugisandLamers:2000} for Wolf-Rayet winds.

Opacities are calculated using OPAL \citep{Iglesias1993,Iglesias1996} Type I and Type II (at the end of hydrogen burning) using \citet{Asplund2009} photospheric abundances. 
Electron screening is included for both the weak and strong regimes using the "extended" option in MESA, which computes the screening factors by extending the classic \citet{Graboske1973} method with that of \citet{Alastuey1978}, and adopting plasma parameters from \citet{Itoh1979} for strong screening. Boundary conditions at the surface of the star are calculated using the \verb|"Eddington_grey"| option of MESA, which uses the Eddington T-tau integration \citep{Eddington1926} to obtain temperature and pressure in the outer layer of the stars. 

Nuclear reaction networks can have a nontrivial effect on the evolution of massive stars \citep{Farmer:2016}, especially the inclusion of elements like iron and nickel \citep{Nabi2019}.
Hence, we make use of an extensive grid of isotopes for the nuclear reaction network in our models to closely follow the evolution of massive stars. The network has been chosen to match globular cluster observations and has 72 elements, including Mg, Li and Fe.
The reaction rates are determined using the Jina Reaclib database \citep{Cyburt2010}.


We use high spatial resolution with \verb|mesh_delta_coeff| = 0.5 and the maximum relative cell size \verb|max_dq| = \timestento{5}{-4}.
Model-to-model structure variation is kept modest with \verb|varcontrol| between \timestento{7}{-4} and \tento{-3}, and additional constraints are used to limit time steps where necessary. 

We compute the evolution of stars in the mass range of 10--110\Msun{} at metallcity, $Z=0.00142$ ([Fe/H] = -1).
All stars start from the pre-main-sequence (PMS) with uniform composition and the goal is to evolve each model until the point of carbon depletion in the core $(X_c\leq10^{-2})$. 
Following \citet{Choi:2016MIST}, we adopt solar scaled abundances from \citet{Asplund2009}, 
with Solar metallicity $Z_\odot= Z_{\odot,\rm protosolar}=0.0142$. The primordial helium abundance $Y_{p}$ is taken to be 0.249 while the protosolar helium abundance $Y_{\odot,\rm protosolar}$ = 0.2703.
Initial hydrogen (X) and helium abundances (Y) for PMS models are calculated using the following formula

\begin{align}
    Y &= Y_{p} + \left(\frac{Y_{\odot,\rm protosolar}-Y_{\mathrm{p}}}{Z_{\odot,\rm protosolar }}\right) Z, 
    \\
    X &= 1 - Y - Z.
\end{align}

Models are evolved without mass loss from the PMS until the zero-age main sequence (ZAMS), defined as when the central hydrogen abundance reduces by 1 percent of the initial value. Evolution is then restarted (including mass loss) using the stellar model saved at ZAMS, and continues until either the termination condition is reached $(X_c\leq10^{-2})$ or the end of 96 CPU-hrs (running parallel on four\LEt{ Write out numerals when lower than eleven and not directly used as a measurement with the unit following. See Sect 2.7 of the language guide https://www.aanda.org/for-authors/language-editing/2-main-guidelines
.} cores for 24 hrs).

The set of models evolved using the physical inputs described above fail to reach the end of carbon burning in the core within the allocated time (24 hrs) for stars more massive than 20\Msun{}. Hereafter, we refer to this set as the "standard set" and label it as "NoModifier" in the figures, since the models were computed without any modifications to facilitate their evolution.
The results for the standard set are unaffected by the increase in temporal and spatial resolution.

\subsection{Results}
\label{sec:results} 

Figure~\ref{fig:Hr_nomodifier} presents the stellar tracks in the Hertzsprung-Russell (HR) diagram for our standard set of models. 
In the left panel, the tracks are colored by the maximum of ${L_{\rm rad}}/{L_{\rm Edd}}$, while the right panel shows the minimum of ${P_{\rm gas}}/{P_{\rm total}}$ for each stellar model.
We see that in these models ${L_{\rm rad}}/{L_{\rm Edd}}$ approaches unity and can even exceed it by factors of a few during the evolution as the stars encounter opacity bumps in their envelopes. These opacity bumps are due to the partial ionization states of iron and helium, as well as the recombination of hydrogen. However, for 10 and 20\Msun{} stars, ${P_{\rm gas}}/{P_{\rm total}}$ remains high enough (>0.5) for their evolution to proceed uninterrupted (see Section~\ref{sec:physics_density_inv}).

Models with initial masses between 30 and 110\Msun{} fail during core helium burning at similar effective temperatures, ${\rm log T_{eff}/K \approx 3.7}$. 
In these models, the stellar envelope inflates in response to the iron-opacity peak, thereby preventing density inversion. However, as the star expands and cools, the minima in gas pressure fraction at the opacity peak decreases as highlighted in the right panel of Figure~\ref{fig:Hr_nomodifier}. The evolution of these models progresses smoothly until stars encounter the opacity peak due to hydrogen and helium in their subsurface layers at ${\rm log T_{eff}/K \approx 4.0}$. The envelope inflation there is not sufficient to prevent the radiative luminosity from exceeding the Eddington luminosity, which leads to density inversions. 

\begin{figure}
    \centering
    \includegraphics[width=\columnwidth]{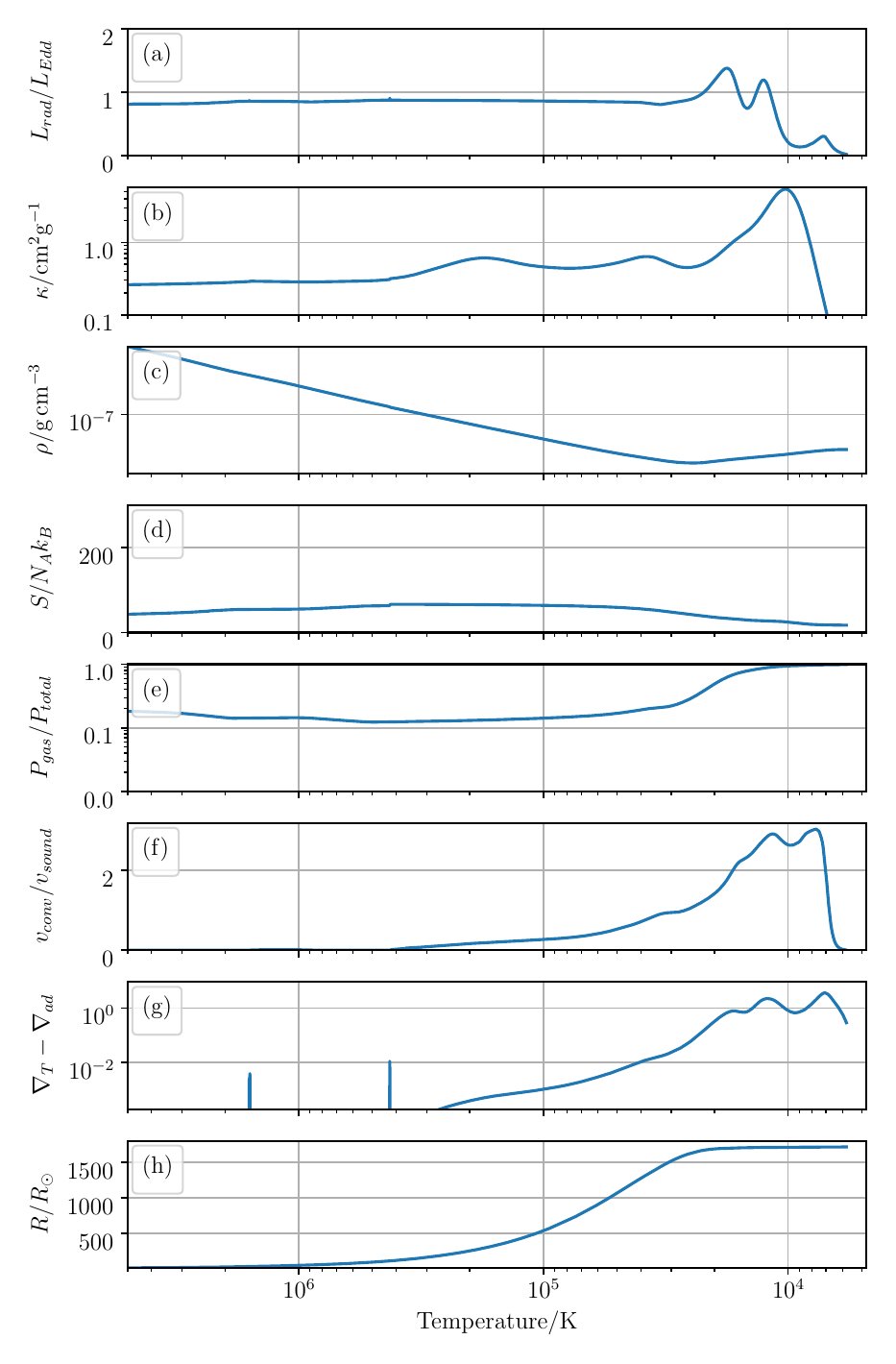}
    \caption{Temperature profile of \revision{the outermost layers of a} 110\Msun{} star computed with the mixing length parameter, $\alpha_{\rm MLT}$ = 5.46. Similar to Figure~\ref{fig:profile_nomodifier_4311}, a density inversion can be seen in panel (c). However, higher convective velocity (panel f) and smaller superadiabaticity (panel g) resulting from the higher value of $\alpha_{\rm MLT}$ helps to keep the specific entropy small (panel d) and time steps large enough to evolve the model without numerical instabilities.}
    \label{fig:profile_mlt}
\end{figure}

To elaborate on the conditions in the stellar interior, the temperature profiles for a 110\Msun{} star at the ZAMS (at ${\rm log T_{eff}/K = 4.77}$), 
and at the end of the track (at ${\rm log T_{eff}/K = 3.72}$, corresponding to the final model reached after 24 hrs of computation), are shown in  Figure~\ref{fig:profile_nomodifier_513} and
Figure~\ref{fig:profile_nomodifier_4311}. \revision{In both these Figures (as well as in Figure~\ref{fig:profile_mlt}, Figure~\ref{fig:profile_ml}, and Figure~\ref{fig:profile_mltpp}) the x-axis is limited to temperatures less than \timestento{5}{6} to show the impact of the various opacity bumps on the evolution of the stellar model).}

In panel (a) of Figure~\ref{fig:profile_nomodifier_513}, \revision{${L_{\rm rad}}/{L_{\rm Edd}}$} shows a small increase corresponding to the opacity peaks due to partial ionization of iron at \timestento{1.5}{6} and \timestento{1.8}{5}\,K (panel b). 
In response to the increase in \revision{${L_{\rm rad}}/{L_{\rm Edd}}$}, gas pressure dips a little (visible as a minimum in ${P_{\rm gas}}/{P_{\rm total}}$ in panel e), although, density consistently decreases outward (shown in panel c). 
The bottom panel (g) shows the difference between the actual temperature gradient and the adiabatic temperature gradient inside the star. 
This difference, ${\rm \nabla_T}-{\rm \nabla_{ad}}$, is known as superadiabaticity (see Section~\ref{sec:mlt++} for details). 
At the location of the opacity peak, superadiabaticity is positive but small and the convective velocity $(v_{\rm conv})$ is less than the isothermal sound velocity $(v_{\rm sound})$ (panel f), signifying efficient convection.
The specific entropy $(S/N_Ak_B)$ in panel (d) at the base of the convective region is small and the evolution of the star proceeds smoothly.

In panel (b) of Figure~\ref{fig:profile_nomodifier_4311}, in addition to the iron opacity peak, opacity peaks corresponding to partial ionization of helium and hydrogen recombination can be seen at \timestento{3.5}{4} and \tento{4}\,K. 
\revision{In contrast to Figure~\ref{fig:profile_nomodifier_513}, the increase in opacity due to hydrogen and helium ionization increases ${L_{\rm rad}}/{L_{\rm Edd}}$ above one, causing the density inversion near the surface of the star.
However, the high value of superadiabaticity (${\rm \nabla_T}- {\rm \nabla_{ad}} \approx 10 $) renders convection inefficient and prone to radiative losses.
Consequently, the specific entropy shows a steep increase  ($S/N_Ak_B\geq300$) while the gas pressure fraction becomes very low $({P_{\rm gas}}/{P_{\rm total}}\lesssim 0.01)$ at the base of the convective envelope near the iron opacity bump.
The convective velocity increases to increase the amount of flux convection can carry, becoming more than the local sound speed. Nonetheless, convection remains inefficient and the high value of entropy causes issues with the convergence of the models \citep{Schwab2019}. Struggling to find a solution, MESA retries with smaller time steps until they become on the order of days}.

The evolution of the star is essentially halted until it is able to get rid of the density inversion as mass loss slowly chips away the outer convective layers. However, evolving the star this way requires a lot of computational time (e.g., it takes $\approx$ 200 CPU-hrs for a 40\Msun{} star) which might not be feasible.

\section{Stellar models: The pragmatic solutions}
\label{sec:model_variations}

\begin{figure}
    \centering
    \begin{tabular}{c}
    \vspace{-0.16in}
\includegraphics[width=\columnwidth]{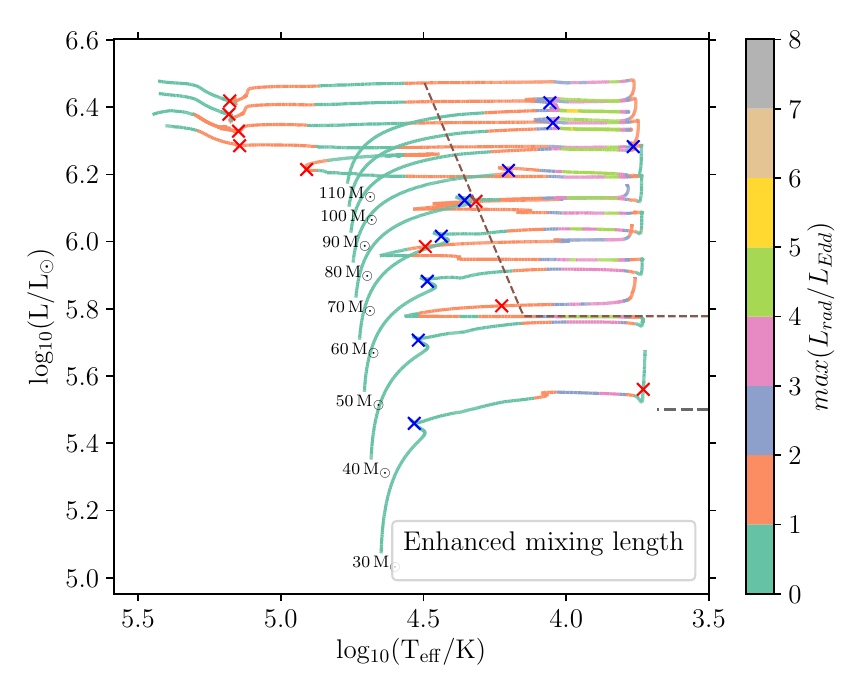}
    \\
    \vspace{-0.16in}
    \includegraphics[width=\columnwidth]{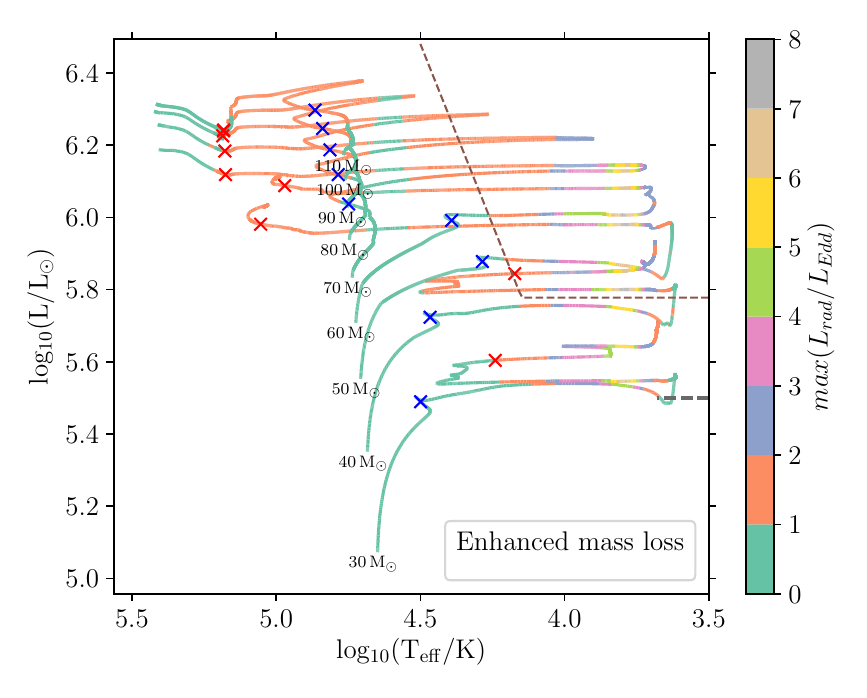}
    \\
    \includegraphics[width=\columnwidth]{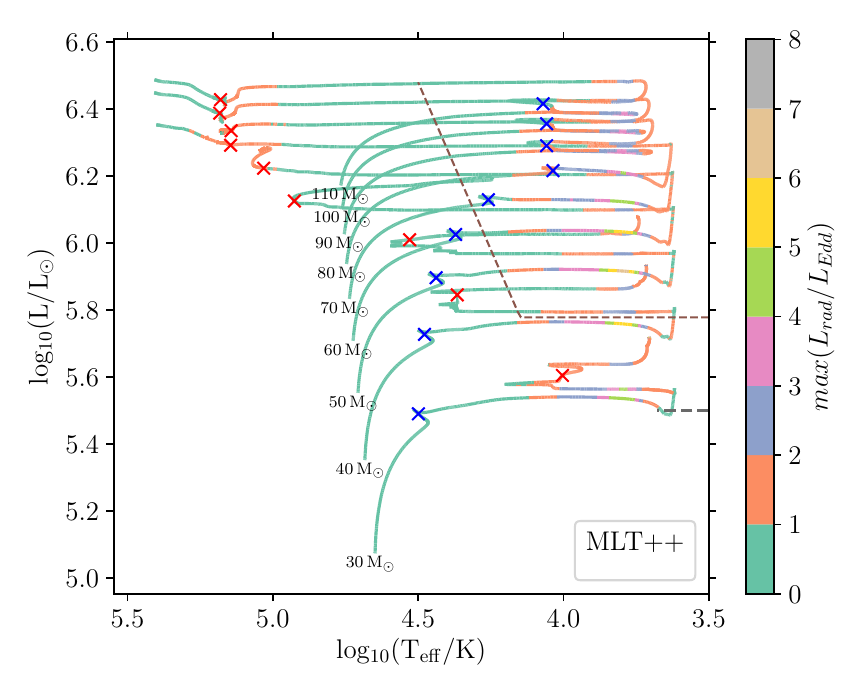}
    \end{tabular}
   
    \caption{HR diagrams showing the evolutionary tracks graded with the maximum of ${L_{\rm rad}}/{L_{\rm Edd}}$ for stars in the mass range 30--110\Msun{}, computed with the model variations described in Section~\ref{sec:model_variations}.
    The top panel shows the tracks computed with enhanced mixing ($\alpha_{\rm MLT}$ = 5.46), the middle panel shows tracks with enhanced mass loss $({\rm \eta_{Dutch}=8.0}$ whenever $L$ exceeds ${\rm L_{Edd}}$) and the bottom panel shows tracks computed with MLT++. Similar to Figure~\ref{fig:Hr_nomodifier}, blue and red crosses mark the end of core hydrogen burning and core helium burning while the brown and gray dashed lines signify the \citet{Humphreys:1979} limit and the \citet{Davies:2018} limit respectively.}
    
    \label{fig:HR_lrad_ledd}
\end{figure}

In the absence of efficient convection, the radiation-dominated envelopes of massive stars in 1D modeling are prone to numerical difficulties \citep{Stothers:1979,Maeder1987}.
Therefore, stellar evolution codes adopt various pragmatic solutions to compute the evolution of massive stars beyond these numerically difficult points \citep{Alongi1993,Ekstroem:2012,Paxton2013}.
The exact solution differs from code to code; however, they can be summed up into three main categories: 
using higher mixing length to increase the efficiency of the mixing process, using higher mass-loss rates to remove layers with numerical instabilities, or suppressing the numerical instability by limiting the temperature gradient and thereby suppressing density inversions.
We explore each of them in detail in the following subsections.

\subsection{\revision{Enhancing mixing length}}

In the convective transport of energy by MLT, the mixing length, $l$
travelled by a fluid element before dissolving in the surroundings is given in terms of the local pressure scale height $H_p$ and mixing length parameter $\alpha_{\rm MLT}$ such that $l = H_p \times \alpha_{\rm MLT}$. 
\footnote{$\alpha_{\rm MLT}$ is a free parameter with a value that is often calibrated from the observations of the sun and eclipsing binaries.  
For example, in the standard set of models in this work the value of $\alpha_{\rm MLT}$ has been calculated using solar data \citep[see][for details]{Choi:2016MIST}}. 
Therefore a higher value of $l$ implies better convective transport of energy and can reduce the value of the Eddington factor, thereby preventing numerical instabilities.

In the low-density envelopes of massive stars, density scale height, $H_{\rho}$, is much greater than the pressure scale height, $H_p$ . Thus, using $H_{\rho}$ instead of $H_p$ in calculating mixing length results in larger $l$ and can help overcome density inversion in the stellar models \citep[e.g.,][]{Ergma1971, Ekstroem:2011}.
Similarly, for a given $H_p$, a higher value of $\alpha_{\rm MLT}$ also implies a higher $l$ and again better convective efficiency. We use the latter approach in this work.
Beginning with $\alpha_{\rm MLT} = 1.82$ (used in the standard set), we compute a series of stellar models with $\alpha_{\rm MLT}=3.0,3.64,4.0,5.0,5.46,7.28,8.0$ for stars with initial mass 30\Msun{} and above. 

We find that for $\alpha_{\rm MLT}\geq$ 5.46, which is three times the value used in the standard set, the models are able to evolve without any numerical instabilities until carbon depletion in the core.

The top panel of Figure~\ref{fig:HR_lrad_ledd} shows the evolutionary tracks evolved with $\alpha_{\rm MLT} = $ 5.46. 
From the figure we see that \revision{${L_{\rm rad}}/{L_{\rm Edd}}$} still exceeds one in the stellar envelope, however, this does not limit the time steps of the computation of the stellar models. The reason for this can be understood from Figure~\ref{fig:profile_mlt}, where we show the temperature profile of 110\Msun{} star at a similar location in the HR diagram where the computation for the 110\Msun{} star from the standard set became stuck.
Similar to Figure~\ref{fig:profile_nomodifier_4311}, the stellar profile contains opacity peaks due to ionization of iron, helium and hydrogen (panel b), resulting in excess \revision{${L_{\rm rad}}/{L_{\rm Edd}}$} (panel a) and the density and gas pressure inversions near the surface (panel c and e). 
However, higher convective velocities resulting from higher $\alpha_{\rm MLT}$ imply that the convective fluid element travels faster and transports more energy before it leaks out due to radiative losses. The superadiabaticity in the outer layers is also smaller compared to the standard case (${\rm \nabla_T}- {\rm \nabla_{ad}} \approx 2 $), meaning radiative losses are also smaller.
The overall convective flux is, therefore, higher than the standard case. The gas pressure fraction is non-negligible $({P_{\rm gas}}/{P_{\rm total}} > 0.01)$ and the specific entropy is small ($S/N_Ak_B<100$). Thus the time steps remain large enough to efficiently compute the evolution of the star until the end of carbon burning.

Increasing convective efficiency in this way helps compute the evolution of stars until core carbon burning. However, it also changes the effective temperature of these stars and makes them appear bluer in the HR diagram \citep{Maeder1987,Kippenhahn2012}.
Comparing the tracks with \revision{enhanced mixing length} (top panel of Figure~\ref{fig:HR_lrad_ledd}), with the standard set (left panel of Figure~\ref{fig:Hr_nomodifier}), we find that stellar models with \revision{enhanced mixing length} are indeed limited to ${\rm log T_{eff}/K \approx 3.73}$ in the HR diagram which is greater than the minimum ${\rm log T_{eff}/K \approx 3.63}$ reached by models in the standard set. We discuss this further in Section~\ref{sec:implications}.

\subsection{Enhancing mass loss}
\label{sec:enhance_mass_loss}

\begin{figure}
    \centering
    \includegraphics[width=\columnwidth]{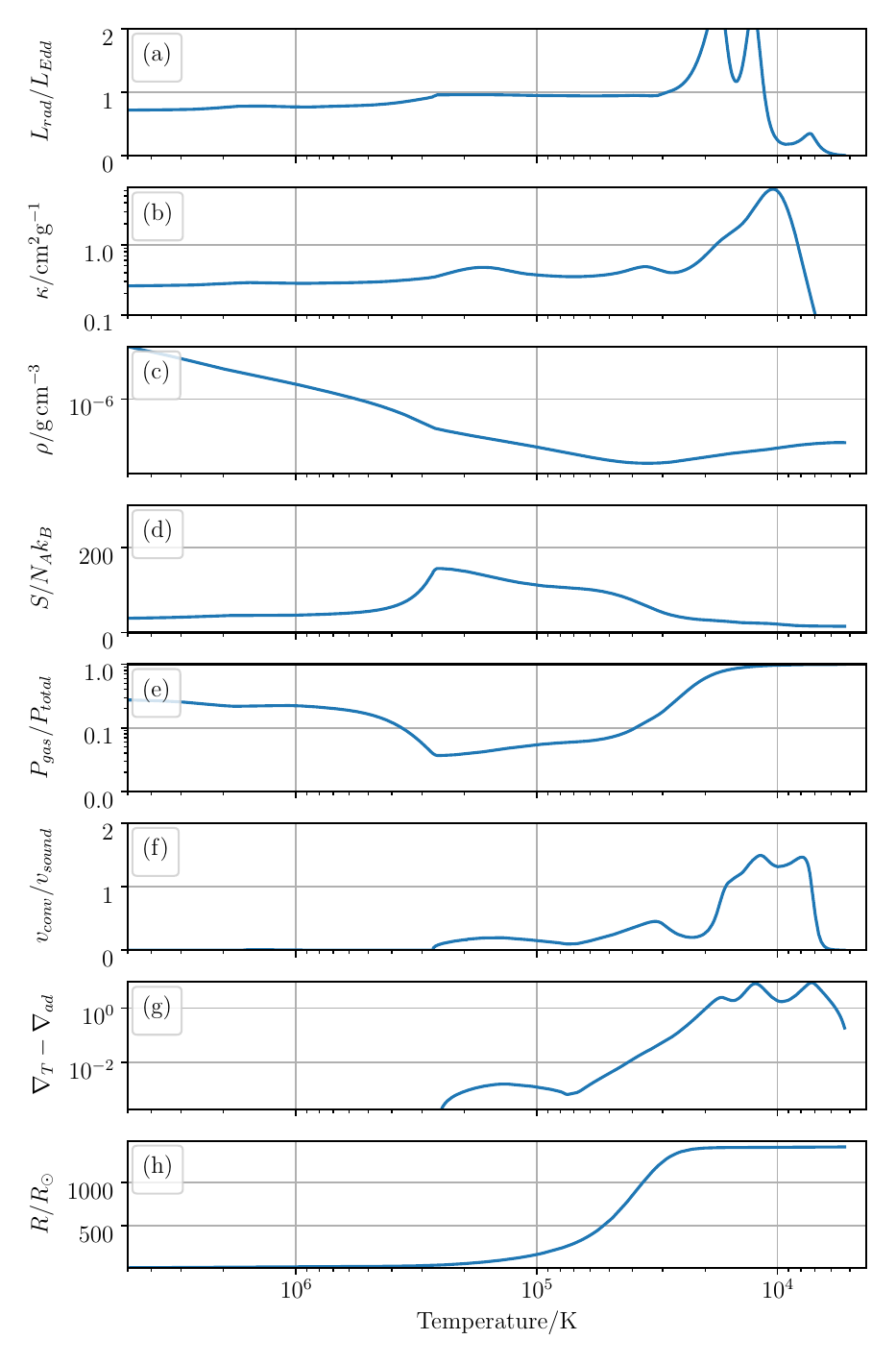}
    \caption{Temperature profile of \revision{the outermost layers of a} 70\Msun{} star computed with \revision{enhanced mass loss}, as described in Section~\ref{sec:enhance_mass_loss}.
    Opacity peaks due to the ionization states of hydrogen, helium and iron and associated density inversions can be seen in panel (b) and panel (c) respectively. However, high mass-loss rates remove the outer layers containing the \revision{hydrogen and helium opacity peaks}, moderating the specific entropy at the base of the convective envelope and the model is able to evolve until completion with reasonably large time steps.}
    \label{fig:profile_ml}
\end{figure}

Winds of hot massive main-sequence stars are radiation-driven winds. Due to their high luminosity, massive stars can generate a high number of photons which can be scattered via ions, transferring momentum with them which then accelerates material outward that can escape the gravitational potential of the star. 
As the star evolves to the cool red-supergiant phase, these winds transition to being dust-driven where they are generated by the interaction of photons with dust grains instead of ions. Massive stars can lose a substantial amount of mass through stellar winds, even their entire hydrogen envelope to become naked helium stars. The mass loss in naked helium stars is again driven by radiation pressure and can be higher by a factor of ten or more than their hydrogen-rich counterparts \citep{Smith:2015}


\revision{Although the mass encompassed in the outer layers of the star which contains the density inversions is small ($<$ 1\Msun{}), the classical mass-loss rates described above are unable to get rid of it as the convective envelope quickly expands to lower temperatures at which hydrogen and helium opacity peaks are present, and density inversions are thereby sustained. However, if the mass-loss rates are high enough, convection will not be able to replenish the envelope and the star will contract and move to higher effective temperatures. High mass-loss rates can therefore help remove outer layers in the stellar model containing the density inversion, helping the stellar model avoid numerical instabilities \citep{Petrovic:2006,Cantiello2009}.}

The mass-loss rates for massive stars are highly uncertain \citep{Renzo:2017,Bjorklund:2021} and can be affected by instabilities and processes other than those described at the beginning of this section. For example, \citet{Grafener:2008} and \citet{Vink:2011} have shown the contribution of optically thick (clumped) winds in the presence of subsurface opacity bumps in massive stars. Moreover, massive stars exceeding the classical Eddington limit can also experience episodes of much stronger mass-loss rates (up to \tento{-3}\Msunyr{}), known as LBV eruptions or super-Eddington winds \citep{Lamers1988, HumphreysDavidson1994,Smith:2004}. 
Despite being the stronger contributor to mass-loss rates for massive stars, the exact rates and the mechanism behind the LBV eruptions remain disputed \citep{Puls2008a, Smith:2014, Owocki2015} and therefore most stellar evolution models often exclude their contributions in the mass-loss rates. 

To determine the impact of enhanced wind mass  loss on the convergence properties of models that fail to evolve in the standard set, we recomputed these models 
with increased wind mass-loss rates for different evolutionary phases until the models are able to evolve without numerical difficulties.
We find that setting the mass loss scaling factor to ${\rm \eta_{Dutch}\geq8.0}$, that is, at least eight times the mass loss used in the standard set,\LEt{ Please avoid the use of italics for emphasis or for words from foreign languages. Please check for this throughout the paper. See Sect. 2.9 of the language guide for more details.}\action{corrected}  whenever the stellar luminosity exceeds the local Eddington luminosity in the envelope, leads to smooth evolution of the stellar models with initial mass greater than or equal to 30\Msun{}. 

We also find that this mass-loss enhancement is only required for stars with a hydrogen-rich envelope (${\rm X_{surf}\geq0.4}$). Although naked helium stars can also exceed the Eddington limit and can develop density inversions in their envelopes, their evolution proceeds uninterrupted for our models without any numerical instabilities. 

\revision{The models with enhanced mass loss as described above, with ${\rm \eta_{Dutch}=8.0}$, are shown in the middle panel of Figure~\ref{fig:HR_lrad_ledd}. A 110\Msun{} star with enhanced mass loss is able to evolve to the end of carbon burning, while the maximum of ${L_{\rm rad}}/{L_{\rm Edd}}$ remains close to unity throughout the evolution. 
Here, the high mass-loss rates and subsequent changes in surface composition due to the removal of H-rich layers reduce the inflation effect \citep{Grafener:2021} and cause the star to lose its envelope before the steep opacity peaks due to hydrogen and helium can develop. 
However, for models less massive than 70\Msun{}, the maximum of ${L_{\rm rad}}/{L_{\rm Edd}}$ can be up to 8, similar to models in the standard set. 
The high values of ${L_{\rm rad}}/{L_{\rm Edd}}$ do lead to the formation of density inversions, as shown in the temperature profile of the 70\Msun{} star in Figure~\ref{fig:profile_ml}. The enhanced mass-loss rates (up to \tento{-4}\Msunyr{}) are unable to prevent the envelope inflation and the models do encounter hydrogen and helium opacity peaks in the cooler outer layers. However, the mass-loss rates are high enough to remove these outer layers containing density inversions before the specific entropy at the base of the envelope becomes too large, or the time steps become too small. Thus, all models are able to evolve smoothly to completion without any difficulty.
}

Enhancing mass loss this way helps compute the evolution of massive stars all the way through to carbon depletion in their core. However, it also influences the structure and evolutionary properties of the stars as explained in Section~\ref{sec:implications}.
\revision{As we discuss further in Section~\ref{sec:observations_hd_limit}, such enhanced mass-loss rates are in tension with recent observational and theoretical estimates for massive stars \citep[e.g.,][]{Bjorklund:2021,Hawcroft:2021A&A}, but may be interpreted as being due to eruptive LBV mass loss\LEt{ Please make sure not to hyphenate "mass loss" when it's being used as a noun; however, it is hyphenated when used as a modifier (e.g., "mass-loss rate").} \citep[e.g.,][]{Smith:2006ApJL,Groh:2020}.}

\subsection{Suppressing density inversions}
\label{sec:mlt++}
\begin{figure}
    \centering
    \includegraphics[width=\columnwidth]{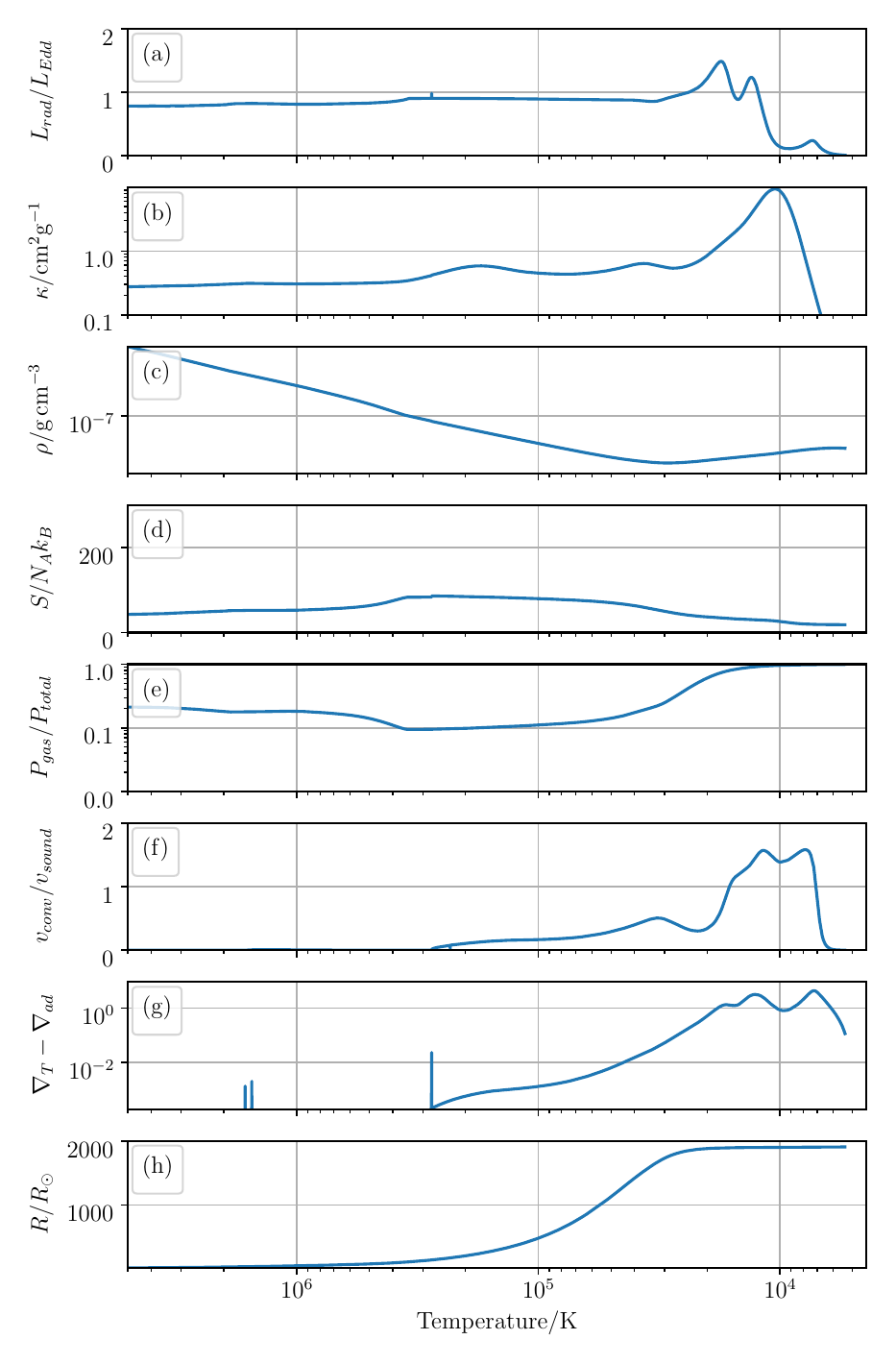}
    \caption{Temperature profile of \revision{the outermost layers of a} 110\Msun{} star computed 
    with the MLT++ method of MESA. The method artificially reduces the difference between ${\rm \nabla_T}$ and ${\rm \nabla_{ad}}$ in the stellar envelope, acting as an source of additional envelope mixing (see Section~\ref{sec:mlt++} for details). The small envelope mixing used here is not enough to suppress density inversions completely (panel c) but it does help to keep the specific entropy small (panel d), gas pressure fraction non-negligible (panel e) and therefore time steps reasonable.}
    \label{fig:profile_mltpp}
\end{figure}

According to the MLT, convection sets in when the temperature gradient of the surrounding material is greater than the gradient interior to the moving element: ${\rm \nabla_T}>{\rm \nabla_{element}}$.
For convection to transport the maximum possible energy, the element should move adiabatically, that is without dissipating energy in the surroundings or ${\rm \nabla_T}\sim{\rm \nabla_{ad}}$.

The difference between ${\rm \nabla_T}$ and ${\rm \nabla_{ad}}$ is termed as the superadiabaticity and is a measure of the efficiency of convection \citep[][]{Maeder2009, Kippenhahn2012}. 
For efficient convection, superadiabaticity is positive but close to 0. It means that the element loses hardly any energy as it traverses the mixing length. 
However, a higher value of superadiabaticity ($\gtrsim$ \tento{-2}) implies that the element suffers energy losses as it travels, and by the time it reaches the end of the mixing length it is left with hardly any energy, thereby, rendering the convective transport of energy quite inefficient.

In the envelopes of massive stars superadiabaticity is the order of unity and the convective transport of energy given by MLT is inefficient, providing fertile ground for density inversions to form and be sustained. Some authors even consider density inversions to be nonphysical, possibly an artifact of the MLT and 1D stellar evolution \citep{Ekstroem:2012} while others consider the possible presence of some unknown mixing mechanism that helps stars get rid of density inversions in nature \citep{Paxton2013}.
Either way, since these density inversions are associated with numerical instabilities in the models of massive stars, many stellar evolution codes suppress them either by limiting ${\rm \nabla_T}$ or by reducing ${\rm \nabla_T}$ to make it closer to ${\rm \nabla_{ad}}$. This reduces the superadiabaticity, thereby making convection efficient and helping stellar models overcome numerical instabilities.

MESA uses the latter approach of reducing ${\rm \nabla_T}$ in the stellar envelope through a method known as MLT++ \footnote{\revision{In the latest MESA release, MLT++ has been replaced by a slightly different method that also works by modifying temperature gradients to boost convective efficiency \citep[see][for details]{Jermyn2022}}}. In this method, whenever the superadiabaticity exceeds a predefined threshold \verb|gradT_excess_f1|, MESA decreases ${\rm \nabla_T}$ to make it closer to ${\rm \nabla_{ad}}$. 
The amount of decrease is given by the combination of the parameter
\verb|gradT_excess_f2| which can be defined by the user, and the parameter \verb|gradT_excess_alpha| which is calculated based on the maximum of ${L_{\rm rad}}/{L_{\rm Edd}}$ and the minimum of ${P_{\rm gas}}/{P_{\rm total}}$ (see Appendix\,\ref{subsec:details_MLT++} for details). 
In general, a smaller \verb|gradT_excess_f2| implies a larger reduction in the superadiabaticity and more efficient convective transport of energy.

We find that using the default values of the MLT++ parameters completely suppresses density inversions but it also gives unrealistic values of luminosity for the most massive stars in our set. 
Therefore, we test the models in the standard set with different combinations of parameters in MLT++, as described in Appendix\,\ref{subsec:details_MLT++}.
Compared to the default values of the MLT++ parameters, we find that using a smaller value of \verb|gradT_excess_f2|=\tento{-1}, and therefore a smaller reduction in the superadiabaticity but occurring
more frequently inside the star (with $\lambda_{1} = 0.6$ and $\beta_{1}= 0.05$), is sufficient for the smooth evolution of the models without any numerical instabilities or inaccuracies. 

The stellar models evolved using MLT++ are shown in the bottom panel of Figure~\ref{fig:HR_lrad_ledd}.
The evolutionary paths of models computed with MLT++ are quite similar to models evolved with \revision{enhanced mixing length}. Although, unlike the models with \revision{enhanced mixing length}, models with MLT++ are not limited to ${\rm log T_{eff}/K \approx 3.73}$ and evolve to lower effective temperatures.
The temperature profile of a 110\Msun{} star evolved with MLT++ (Figure~\ref{fig:profile_mltpp}) again shows a similar behavior compared to the temperature profile of the 110\Msun{} star evolved with \revision{enhanced mixing length} (Figure~\ref{fig:profile_mlt}), at similar ${\rm log L}$ and ${\rm log T_{eff}}$. However, MLT++ artificially reduces ${\rm \nabla_T}$, such that
the superadiabaticity and the specific entropy at the base of the convective envelope remain small despite having lower convective velocity compared to the 110\Msun{} model with \revision{enhanced mixing length}.

\section{Impact on the stellar properties}
\label{sec:implications}

Each of the three solutions described in Section~\ref{sec:model_variations} help compute the evolution of massive stellar models in the standard set until the end of carbon burning. However, in the process they also modify the evolutionary pathway of the stars and impact their evolutionary outcome. In this section, we compare the set of stellar models obtained with the minimum numerical enhancement from each solution and determine the impact of these solutions on the structure and evolution of the massive stellar models. 

\subsection{Structure of the star}
\label{sec:structure}
\begin{figure*}
    \centering
    \begin{tabular}{ccc}
    
    \hspace{-0.2in}

    \includegraphics[width=0.34\textwidth]{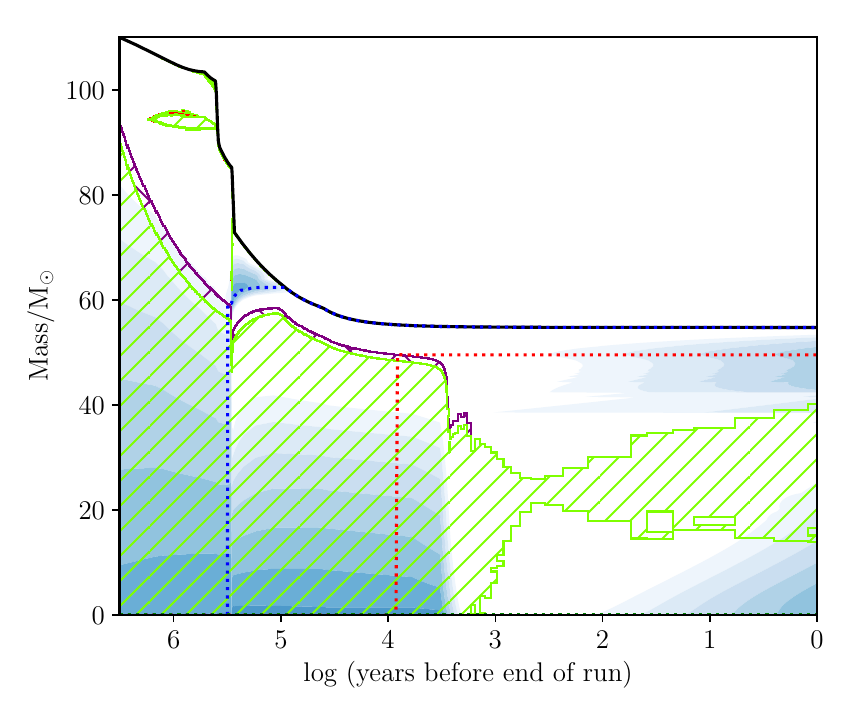}
    
    &
   
    \hspace{-0.2in}

    \includegraphics[width=0.34\textwidth]{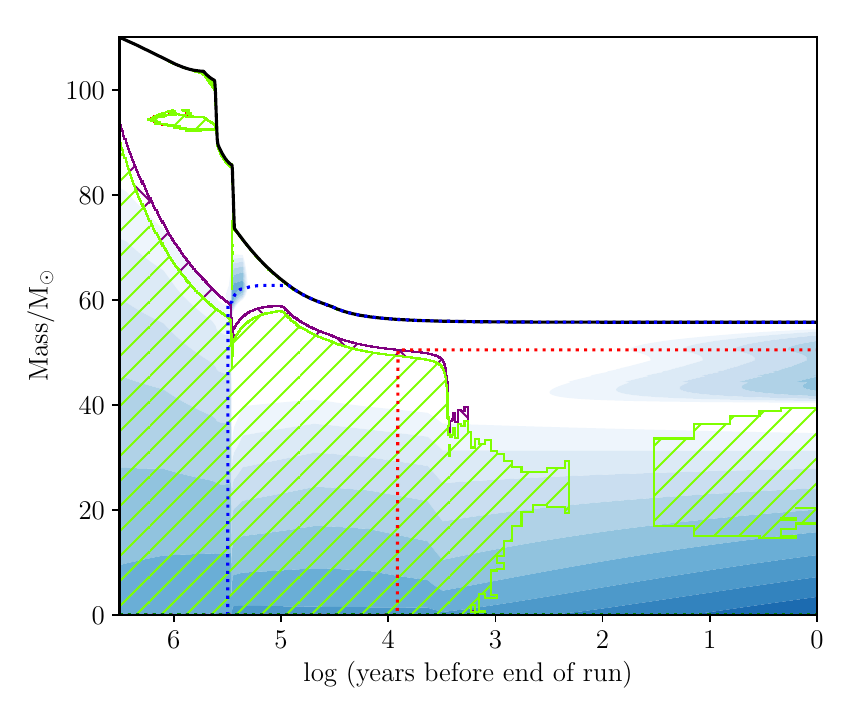}

    &
    
        \hspace{-0.2in}

    \includegraphics[width=0.34\textwidth]{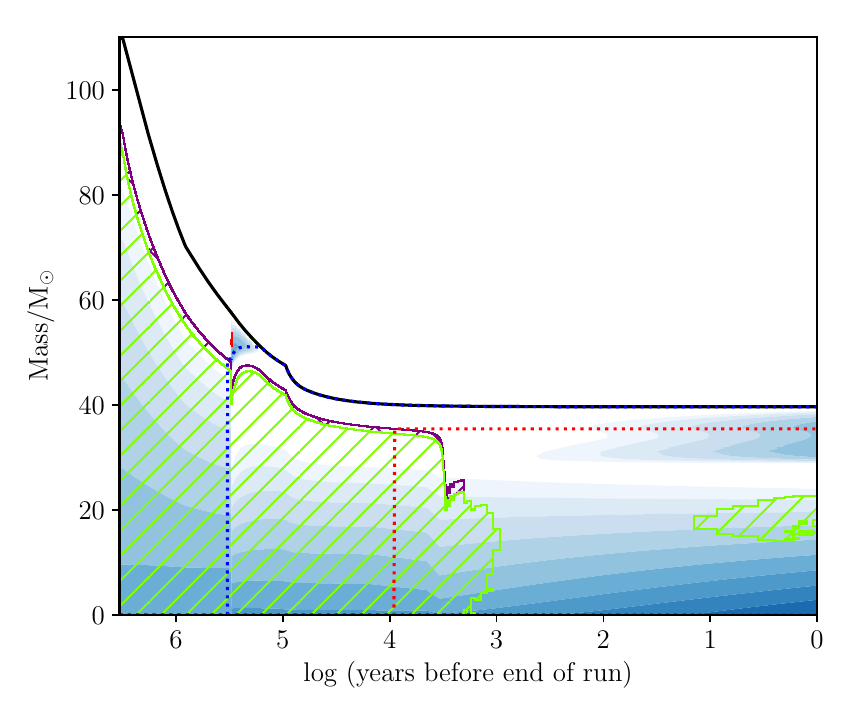}

    \end{tabular}
    \caption{Kippenhahn plots showing the structure of a 110\Msun{} star evolved with enhanced mixing (left), with MLT++ (middle) and with enhanced mass loss (right). The Y-axis represents the mass coordinate inside the star while the X-axis represents the time remaining in the life of the star before the end of the run is reached.
    Green, purple and red hatching mark the regions with convection, overshooting and semiconvective mixing, respectively.  In all three panels, the helium core boundary is the outermost location where the hydrogen mass fraction is $<$ 0.01, while the helium mass fraction is $\geq$ 0.01, and is represented by the blue dashed line. Similarly, the carbon core boundary is defined as the outermost location where the hydrogen and helium mass fraction are $<$ 0.01 while the carbon mass fraction is $\geq$ 0.01. It is represented as the red dashed line. 
    }
    \label{fig:kippenhahn}
\end{figure*}

Figure~\ref{fig:kippenhahn} shows the Kippenhahn plot of a 110\Msun{} star---depicting regions within the star by mass as a function of time in the period leading up to the end of the run---for each of the models computed with the pragmatic solution described in Section~\ref{sec:model_variations}. The evolution of the 110\Msun{} model evolved using MLT++ is similar to the model with \revision{enhanced mixing length} but quite different to the evolution of the model computed with \revision{enhanced mass loss}.

All three models start with a 90\Msun{} convective core, (shown by the green hatching) accompanied by a 3\Msun{} overshoot region outside the core, (shown by the purple hatching). The convective core decreases in size as the star evolves through the main sequence.

In models computed with \revision{enhanced mixing length} and MLT++, thin strips of convection, two close to the surface and the third at about 90\Msun{}, are formed as the star encounters the hydrogen, helium and iron opacity bumps in the final \tento{6} years of its evolution. The envelope inflation due to the iron opacity bump causes the star to become a red supergiant before core hydrogen burning can finish. The star suffers higher mass-loss rates as a red supergiant that can be -- seen as a rapid decline in the total mass of the star (black solid line) -- until a 60\Msun{} helium core is formed -- depicted by a blue dotted line in the figure. The star continues to lose mass, as the core helium burning and the hydrogen shell burning ensues, although at a comparatively lower rate. It ultimately loses its envelope to form a naked helium star in the final \tento{5} years of its evolution, before finally forming a roughly 50\Msun{} carbon core in the end. 

\begin{figure}
    \centering
    \includegraphics[width=\columnwidth]{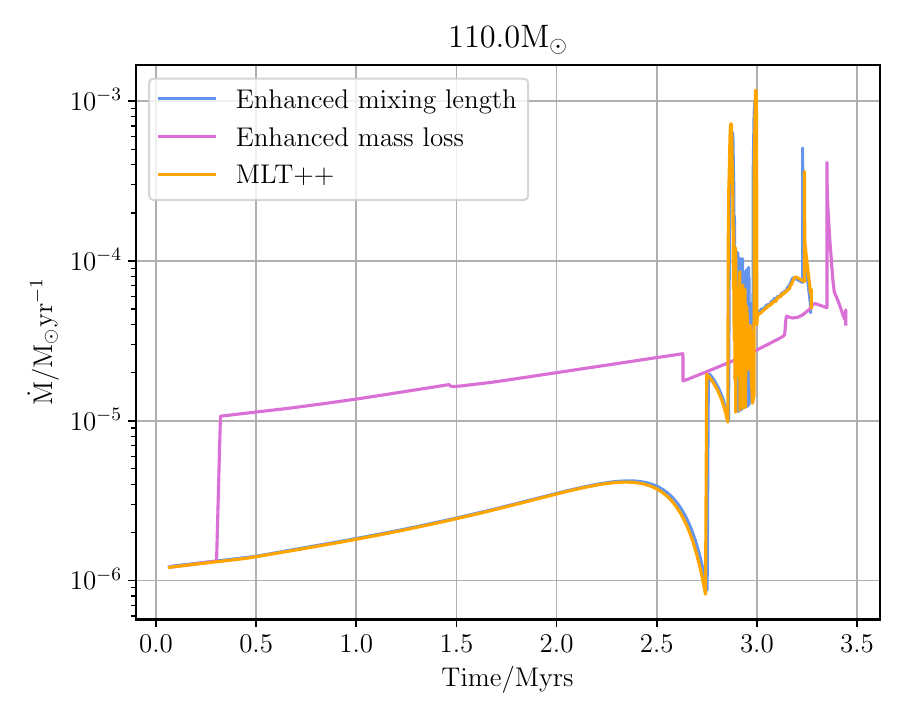}
    \caption{Variation in mass-loss rates with time for a 110\Msun{} star evolved with \revision{enhanced mixing length} (blue line), \revision{enhanced mass loss} (pink line) and with MLT++ (orange line). While the maximum mass-loss rate encountered by the 110\Msun{} model with \revision{enhanced mass loss} is less than the maximum mass-loss rate encountered by the models evolved with the other two solutions, it still ends up  being the least massive of all in the end. See Section~\ref{sec:structure} for an explanation.
    }
    \label{fig:mass_loss_rate}
\end{figure}

In the model computed with enhanced mass loss, mass-loss rates are quite high ($> 10^{-5}$\,\Msunyr) during the main-sequence evolution. 
The convective core shrinks rapidly in response to high mass-loss rates. The star loses its outer layers before density inversion due to hydrogen and helium opacity bumps can happen, becoming a naked helium star shortly after a 50\Msun{} helium core is formed. The final product is a 40\Msun{} naked helium star with a 35\Msun{} carbon core.

The similarities in the evolution of the models using MLT++ and enhanced mixing can be understood as follows. 
In models with enhanced mixing, convection is already efficient in the core, owing to its high density and increasing $\alpha_{\rm MLT}$ (and therefore the mixing length) hardly makes any difference. However, the density in the subsurface of layers of the star can be $\approx$\tento{-10}${\rm \,g\,cm^{-3}}$ and convection is highly inefficient, and therefore increasing $\alpha_{\rm MLT}$ leads to more efficient convective transport of energy in the stellar envelope.  

For models using MLT++ to suppress density inversions, the story is similar. Reducing the temperature gradient (superadiabaticity) prevents radiative losses from the convective cells, making them more efficient at transporting energy. However, near the center convection is nearly adiabatic and the value of superadiabaticity is small (<\tento{-4}), therefore MLT++ is not applicable there. Thus models with high $\alpha_{\rm MLT}$ and MLT++ produce similar core structures.

For models with extra mass loss the evolution is quite different from the first two cases as the star loses quite a lot of mass even on the main sequence. 
This same trend continues for the post-main-sequence evolution. Therefore, it has the least of both helium and carbon core masses. Similar to the other two solutions, the 110\Msun{} model again loses all its envelope and ends up as a naked helium star during the core helium burning phase. 

Interestingly, the maximum mass-loss rate encountered by the 110\Msun{} star with \revision{enhanced mass loss} is lower than the models with enhanced mixing and with MLT++, as shown in Figure~\ref{fig:mass_loss_rate}. However, models with \revision{enhanced mass loss} by construction have higher mass-loss rates during most of the main sequence, therefore they become a naked helium star without ever undergoing the red-supergiant phase where the peak in mass-loss rates usually occurs (due to the large radius of the star).
Models with enhanced mixing and MLT++ lose their envelope later in the evolution as they encounter high mass-loss rates as a red supergiant and therefore end up with higher total mass compared to the model with \revision{enhanced mass loss}.

\subsection{Final mass and remnant properties}
\label{subsec:final_mass_remnants}

\begin{figure}
    \centering
    \begin{tabular}{c}
         \includegraphics[width=\columnwidth]{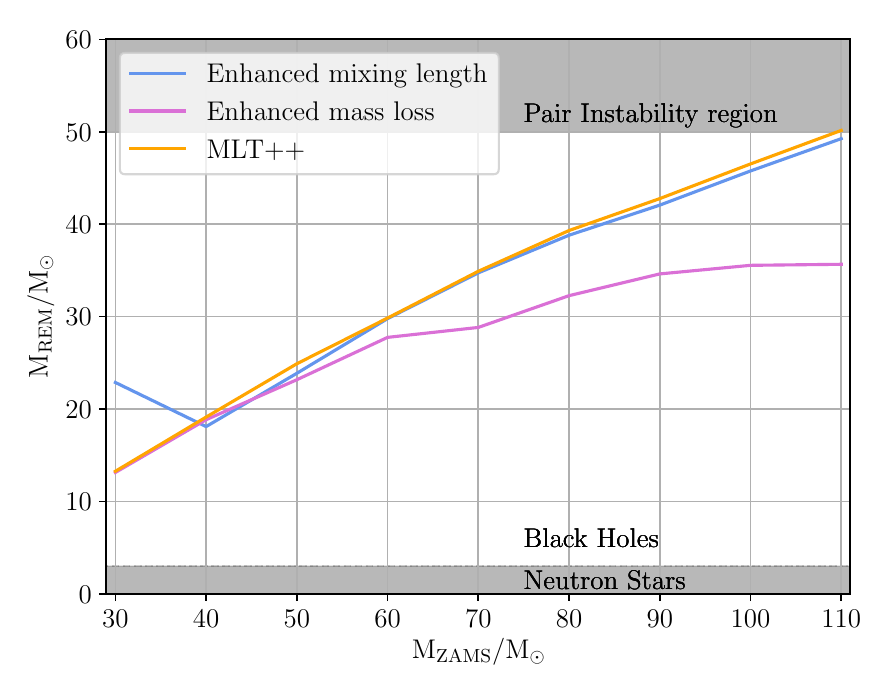}  \\
         \includegraphics[width=\columnwidth]{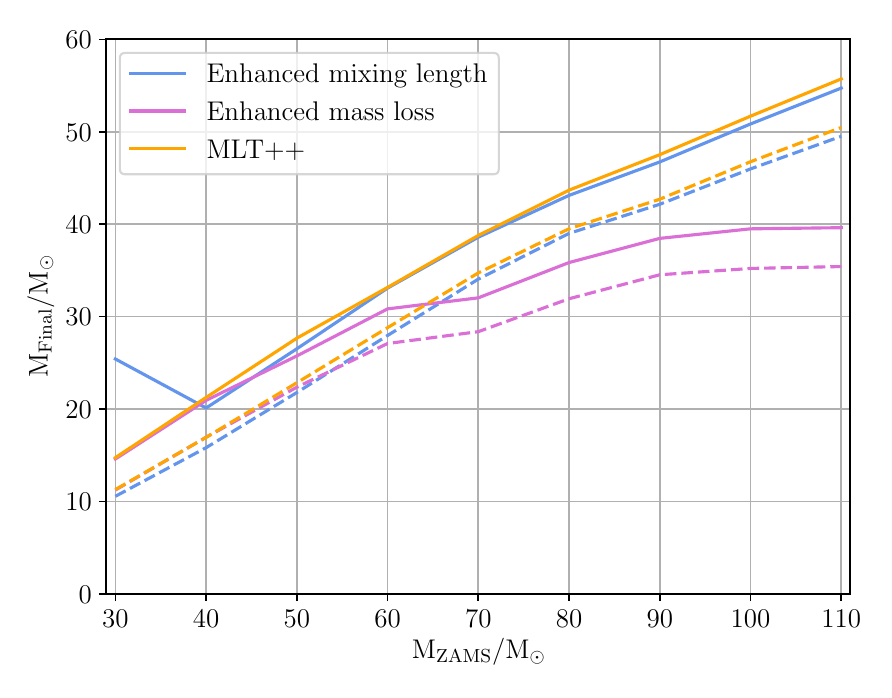}
    \end{tabular}
    
    \caption{\LEt{ If possible, please add a general title without an initial article before  "The top panel" here and anywhere else where one is not provided for figure or table legends.}\action{corrected} \rev{Initial versus final stellar masses for the three sets of stellar models that were computed using the pragmatic solutions.} The top panel shows the mass of stellar remnants as a function of their initial mass, as predicted by the \citet{Belczynski:2008} prescription for the sets of stellar models computed using the pragmatic solutions. The bottom panel shows the final total mass (solid line) and the carbon-oxygen core mass (dashed line) that were used in calculating the remnant mass. For stars more massive than 60\Msun{} the differences in the remnant mass can be up to 14\Msun{}.}
    \label{fig:remnant_mass}
\end{figure}

Massive stars are expected to end their lives in supernovae, leaving behind compact remnants (neutron stars and black holes) in a core-collapse or a pulsational-pair instability supernova or undergoing complete disruption  
in a pair-instability supernova. Their demise as a supernova explosion is also important for modifying the chemical and energy makeup of their surroundings, paving way for the formation of new generations of stars. Furthermore, stellar remnants are important for studies across a wide spectrum. They are the progenitors of X-ray binaries, gamma rays bursts, and compact binary mergers that lead to gravitational waves. Hence, it is important to quantify the differences between the properties of the remnants formed by the massive star models.

There are many prescriptions available in the literature that relate the final properties of the star with the mass of the remnant it would form \citep[e.g.,][]{Eldridge2004,OConnor:2011,Fryer:2012,Ertl:2015}.
Here we use the \citet{Belczynski:2008} prescription, which is the same as the StarTrack prescription in \citet{Fryer:2012}, to calculate the mass of the stellar remnants for each set of models. The method uses the total mass and the core mass of the star at the end of carbon burning to calculate the mass of the remnant \citep[we refer the interested reader to Section 6.1 of][for further details of the method]{Agrawal:2020}.

Following \citet{Belczynski:2010}, we plot the remnant mass of the stars as a function of their initial mass as given by the three sets of stellar models computed using the pragmatic solutions in Figure~\ref{fig:remnant_mass}. 
The top panel of the figure shows the remnant mass of the stars while the bottom panel shows the final total mass of the star and the mass of the carbon-oxygen core. 

For all the sets of models, the \citet{Belczynski:2008} prescription predicts the formation of black holes with masses in the range of 13--50\Msun{}.
The difference in the mass of black holes as predicted by each set is less than 2\Msun{} for stars with initial masses up to 60\Msun{}, with the exception of the 30\Msun{} star where the model with \revision{enhanced mixing length} predicts a significantly higher remnant mass (22\Msun{} black hole) compared to other two sets (13\Msun{} black hole). 
For stars more massive than 60\Msun{} the curve diverges rapidly, and the difference between the remnant masses can be up to 14\Msun{}. A similar trend can be seen in the total mass and core masses for each set. For the 30\Msun{} star, the model with \revision{enhanced mixing length} predicts a higher final total mass but lower carbon-oxygen core mass compared to the other two sets. This is because a larger mixing length in the model leads to a more compact star with lesser mass loss. Thus, the 30\Msun{} model is able to retain most of its envelope and ends up with the final total mass of 26\Msun{}. 
Similarly, the origin of differences in the black hole mass predictions can be traced back to the mass-loss rates experienced by each model which themselves are dependent on the surface properties of the star. 

\revision{In models with enhanced mass loss, stars with initial masses greater than 50\Msun{} lose their hydrogen envelopes due to mass-loss and become Wolf--Rayet stars (cf. Figure~\ref{fig:HR_lrad_ledd}).
As mentioned in Section~\ref{sec:standard_model}, we use the mass-loss prescription from \citet{NugisandLamers:2000} for Wolf--Rayet stars.
Whilst more modern Wolf--Rayet mass-loss prescriptions exist \citep[e.g.,][]{Grafener:2005,Vink:2005}, and may lead to different absolute mass-loss rates, we expect that the \emph{relative} differences between the models shown in Figure~\ref{fig:remnant_mass} should be robust to this difference \citep[see for example][]{Belczynski:2010}.
}

\subsection{The maximum radial expansion}

\begin{figure}
    \centering
    \includegraphics[width=\columnwidth]{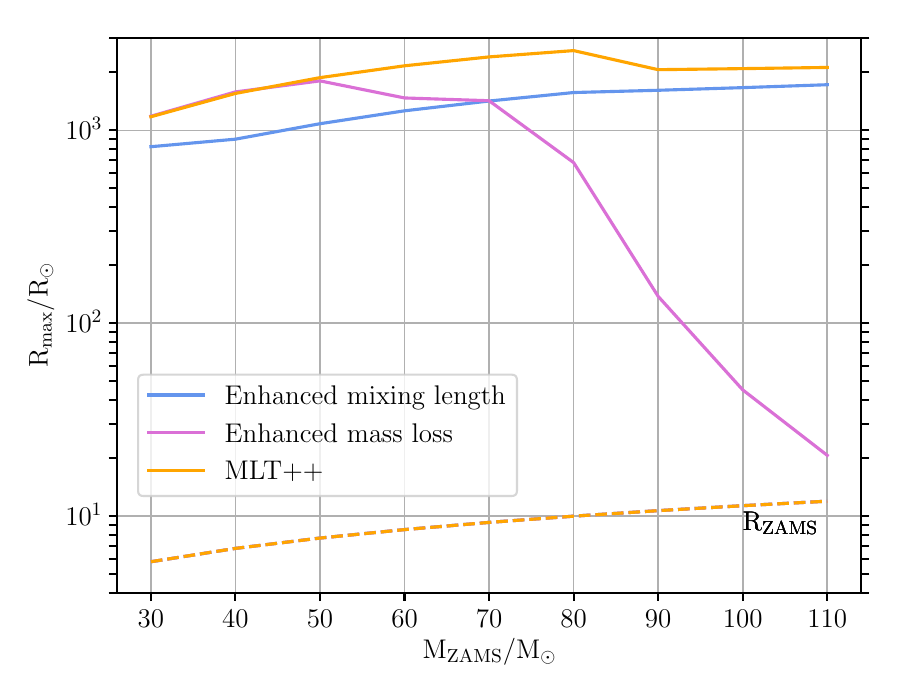}
    \caption{Maximum radial expansion achieved by a star during its lifetime as a function of its initial mass, as predicted by the three sets of stellar models computed using the pragmatic solutions. The difference in predictions of the maximum radial expansion by each model varies between 500--2000\Rsun{}, and can have important implications if the star is in a binary system.}
    \label{fig:maxrad}
\end{figure}

Figure~\ref{fig:maxrad} shows the maximum radial expansion achieved by the stars during their evolution, computed with the three different pragmatic solutions. 
The maximum difference in the maximum radial expansion is $\sim$2000\Rsun{} which occurs between models with \revision{enhanced mass loss} and models with MLT++, for the most massive 110\Msun{} star in the set.  
For models with \revision{enhanced mixing length} and MLT++ which appear to undergo similar evolutionary paths and final fates, the difference in maximum radial expansion can still be up to 1000\Rsun{}, especially for stars in the mass range 30--80\Msun{}. 
Even for the 110\Msun{} star, where the difference between models with \revision{enhanced mixing length} and MLT++ appears to be the least, the maximum radius can differ by 500\Rsun{}.
This has important implications for the binary evolution of the star, as radial proximity determines the episodes of mass transfer in close binary systems. 

These differences in the stellar radii are again the result of the pragmatic solutions used in each set. 
For models with initial masses greater than 60\Msun{} and computed with \revision{enhanced mass loss}, high mass-loss rates strip the envelope of the stars before they can become a red supergiant. Thus these stars evolve directly toward the naked helium star phase and show the least radial expansion. For stars with \revision{enhanced mixing length}, a higher mixing length means the fluid element is more efficient at transporting energy through convection. This decreases the size of the convective cells in the envelope and the model remains more compact and at higher effective temperatures than the models from the other two sets. 

An interesting case is presented by models with MLT++ where reducing the temperature gradient can have the same effect as excess envelope mixing \citep{Sabhahit:2021}. 
For stars up to 50\Msun{}, models using MLT++ closely mimic the behavior of models with \revision{enhanced mass loss}, although beyond 50\Msun{}, the excess envelope mixing in models with MLT++ makes them resemble more closely the models with enhanced mixing. 


\section{Clues from observations: The Humphreys-Davidson limit}

\label{sec:observations_hd_limit}

It is well-established that the evolution of massive stars is highly uncertain. It is riddled with many physical and numerical problems. Therefore the need for pragmatic solutions arises. 
To make matters worse, it is difficult to say which solution is closer to reality as there is an apparent scarcity of massive stars in the region of the HR diagram where numerical issues with density inversions occur.

First characterized by \citet{Humphreys:1979}, the Humphreys-Davidson (HD) limit defines the region in the HR diagram above ${\rm log L / L_\odot = 5.8}$ where very few massive stars in the Galaxy have been observed to date. 
Using recent observations of red supergiants in the Large Magellanic Cloud (LMC) and the Small Magellanic Cloud (SMC), \citet{Davies:2018} found the limiting luminosity of the brightest red supergiants to be
${\rm log L / L_\odot =5.5}$ in both galaxies, slightly lower than previous studies.
The LMC and SMC are lower metallicity environments than the Milky Way, closer to the metallicity used in this work.
The findings of \citet{Davies:2018} also suggest that the maximum red-supergiant luminosity does not vary strongly with metallicity.

While the absence of massive stars as red supergiants beyond the HD limit may not help trace the exact evolutionary path of massive stars, it does provide an important clue that stars more massive than about 40\Msun{} do not spend much time in the HD region. 
In recent years, several studies \citep[e.g.,][]{Castro:2018,Kaiser:2020,Vink2021b} have tried to constrain the different mixing mechanisms (such as semiconvection and overshooting), mass-loss rates, and binary properties of massive stars by using stellar models evolved with different physical inputs to reproduce the HD limit. However, they also employ pragmatic solutions for instabilities due to density inversions to compute the complete evolution of the stellar track. 
Using these pragmatic solutions interferes with other physical inputs and adds an implicit bias in the computation of models.
As these pragmatic solutions can be different across different stellar evolution codes, results obtained with them might not reflect the true value of the physical parameter being constrained for the massive stars. 

Studies that do not employ any pragmatic solutions \citep[e.g.,][]{Klencki:2018} are usually limited to stars less massive than 30\Msun{} or to the evolutionary phases before numerical instabilities arise in more massive stars. While this helps avoid bias due to pragmatic solutions, it also inhibits exploring the late-stage evolution of massive stars, such as the end of core helium burning and beyond. Thereby, affecting the potential studies involving stellar remnants and transients.

Some studies also show that the HD limit can also be reproduced by stellar models by just using these pragmatic solutions.  
For example, using just MLT++ as the source of excess envelope mixing in massive stars up to 50\Msun{}, \citet{Sabhahit:2021} were able to reproduce the lack of massive stellar models beyond the HD limit and the Davies limit at Galactic ($Z=0.017$), LMC ($Z=0.008$) and SMC ($Z=0.004$) metallicity.

\begin{figure}
    \centering
    \includegraphics[width=\columnwidth]{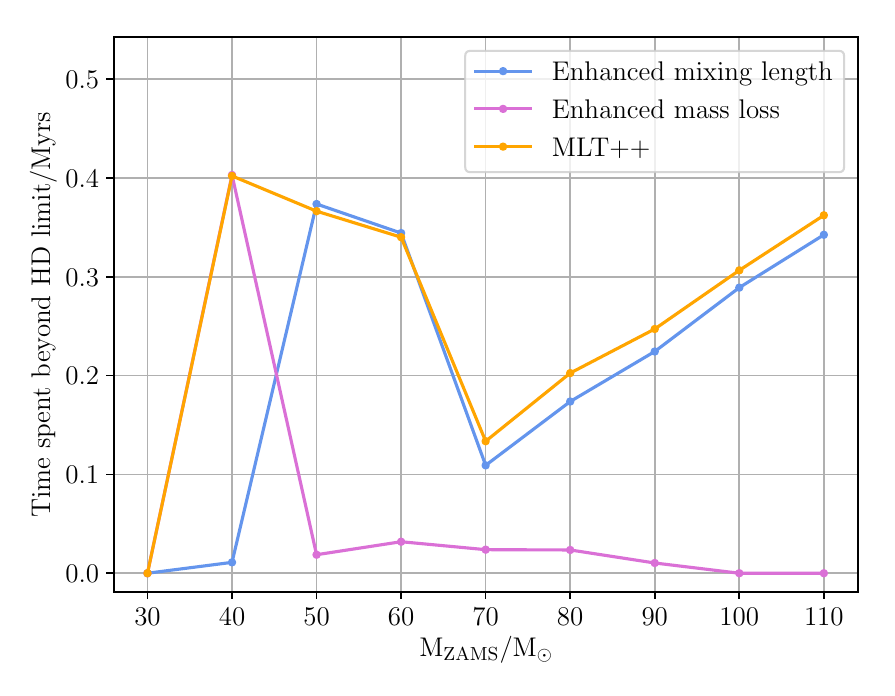}
    \caption{Time spent beyond the HD limit as a function of the initial mass of the star as given by each set of stellar models computed using the pragmatic solutions.}
    \label{fig:hd_time}
    
\end{figure}

Recently, \citet{Gilkis2021} showed that using significantly enhanced mixing parameters in stellar models can reduce the time spent by stars beyond the HD limit.
Following \citet{Gilkis2021}, in Figure~\ref{fig:hd_time} we show the amount of time stars spend beyond the HD limit as a function of initial mass  for each of our sets of models. 
We see that the models computed with \revision{enhanced mass loss} spend the least time (except for a 40\Msun{} stellar model) while most of the massive stars computed with enhanced mixing and MLT++ can spend between \timestento{2}{5} and \timestento{4}{5} years in the HD region. Therefore, models with \revision{enhanced mass loss} may appear to be closest to observations (or lack of observations) of massive stars. This, however, has serious implications for gravitational wave observations, as the maximum black hole mass predicted by this set of the model is just 35\Msun{} (see Figure~\ref{fig:remnant_mass}).

\begin{figure}
    \centering
    \includegraphics[width=\columnwidth]{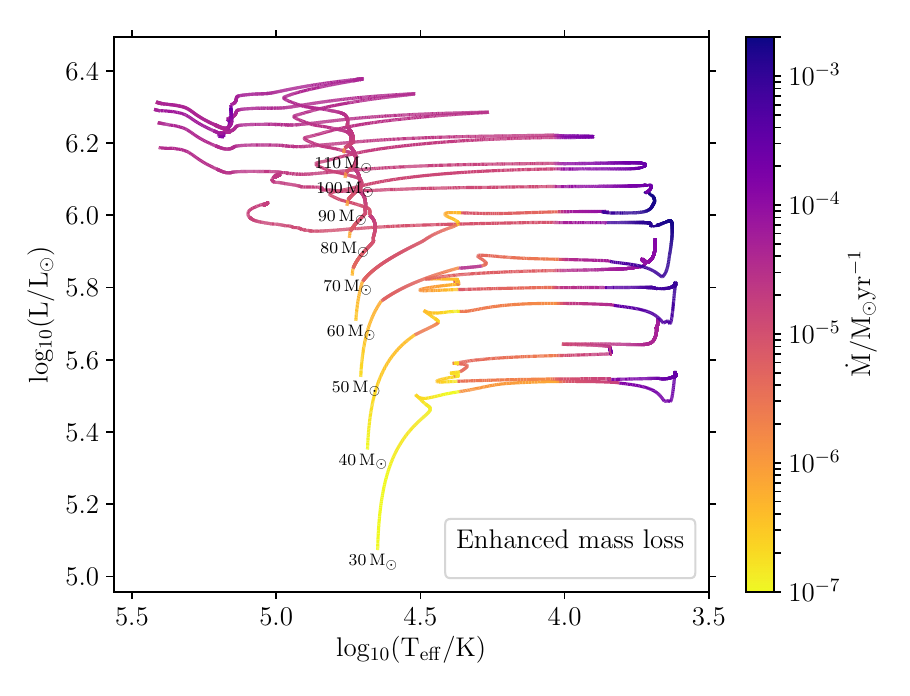}
    \caption{HR diagram showing the mass-loss rates for stars in the mass range 30--110\Msun{}, computed with the \revision{enhanced mass loss} described in Section~\ref{sec:enhance_mass_loss}. 
    }
    \label{fig:hr_mdot_ml}
\end{figure}

To unravel this problem, we plot the evolutionary tracks, colored according to the mass-loss rates, from the set computed with the \revision{enhanced mass loss} in Figure~\ref{fig:hr_mdot_ml}. 
Despite the enhancement near the Eddington limit, the maximum mass-loss rates for models with enhanced mass loss seem to be consistent with the typical mass-loss rates for massive stars \citep[][]{Smith:2014}.
The maximum mass-loss rate of \timestento{1.8}{-3}\Msunyr{} is experienced by stars in the 40--60 \Msun{} mass range
during the red-supergiant phase of their evolution. However, these rates only last for a maximum of \timestento{7}{3}\,yrs. For stars more massive than 60\Msun{}, the maximum mass-loss rates are an order of magnitude lower, about \timestento{3}{-4}\Msunyr{}, and last between \timestento{3}{2}--\timestento{2}{4}\,yrs. The lower mass-loss rates encountered by more massive stars are a consequence of the enhanced mass loss during their main-sequence evolution. These stars experience mass-loss rates of about \tento{-5}\Msunyr{} for more than \timestento{3}{5} years. Therefore, they lose a significant portion of their envelope while on the main sequence and do not undergo the red-supergiant phase of evolution.

Whether massive stars can retain these high mass-loss rates for prolonged periods of time is currently questionable. 
\revision{Recent (theoretical and observed) estimates of steady-state mass-loss from massive O stars ($M \gtrsim 50$\Msun{}) suggest that the mass-loss rates for these stars should be lower, rather than higher, than those typically used in computing single star models \citep[][]{Beasor2020,Vink2021a,Bjorklund:2021,Hawcroft:2021A&A}.
However, these massive stars also experience eruptive mass-loss during LBV phases, which can be important for the mass-loss history of these stars \citep[][]{Smith:2006ApJL}, with effective mass-loss rates up to $10^{-4}$\Msunyr{}. 
Such mass-loss is often neglected in one-dimensional stellar models, but can be approximately included as an average, enhanced mass-loss rate, as we do here \citep[e.g.,][]{Groh:2020, Sabhahit:2022}.}
Interaction with a companion in a binary system can also lead to higher mass-loss rates, but again the smaller radii predicted by these models (Figure~\ref{fig:maxrad}) make binary interactions less likely. 
Therefore, the validity of mass-loss rates in models computed with extra mass-loss, or in general any of the pragmatic solutions used by the codes cannot be ascertained at present.

Mixing processes such as rotation, semiconvection, and overshooting have also been shown to have a large impact on the red-supergiant phase of evolution for massive stars and on the reproducibility of the HD limit  \citep{Schootemeijer:2019,Higgins:2020njv,Gilkis2021,Sabhahit:2021}.
Further, many studies suggest that magnetic fields can have a significant effect on the subsurface convection regions of the stars and hence on the density inversions. For example, \citet{Jiang2018} have shown that the turbulent velocity fields around the iron opacity peak can be escalated in the presence of magnetic fields. 
We have not explored the role of semiconvection, rotation, magnetic fields and binarity in this work, each of which we acknowledge can have a significant impact on the evolution of massive stars.
Evolving massive stellar models with these properties is important and will be a part of future work.

\section{Conclusion}

\label{sec:conclusion}

\rev{In this work, we performed a systematic study of the density inversions arising due to the super-Eddington layers in the envelope of the massive stars and their impact \LEt{ The intended meaning of "inversions on computing" isn't clear, please rephrase, possibly writing "inversions by computing" or "inversions computing".}\action{corrected} on computing the evolution of massive stars with the 1D stellar evolution code MESA. }
Using commonly used input parameters for massive stellar models, we computed the nonrotating evolutionary tracks of 10-110\Msun{} stars for $Z=0.00142$. 
Models with initial masses between 30 and 110\Msun{} in this standard set fail to reach the end of core carbon depletion $(X_c\leq10^{-2})$ as the time steps become too small (on the order of days) as models encounter density inversions in the envelope. We find that this is due to large specific entropy and a low gas pressure fraction at the iron opacity bump and a high Eddington factor at the\LEt{ Please avoid the use of slashes. For details, refer to Sect. 2.9 of the language guide. Please check for this throughout.
Here, you might write "hydrogen-helium" instead if possible.}\action{corrected} hydrogen \rev{and} helium opacity bumps in the stellar envelope. 

We recomputed the models for 30--110\Msun{} stars using three pragmatic solutions: enhancing the mixing length parameter, enhancing mass-loss rates, and suppressing density inversions by adopting MLT++. 
\rev{These three\LEt{ Please add a noun here, possibly "actions" or "solutions".} \action{corrected} solutions are a general form of those used by different stellar evolution codes to resolve numerical instabilities associated with density inversions in models of massive stars. }
With each solution, we were able to compute the evolution of massive stars up to carbon depletion in the core. 

To determine the least possible impact each solution can have on the evolution of the star, we picked the sets with the minimum enhancement for each solution. With these, we find that density inversion regions can still form in these models, but the evolution of models proceeds smoothly, that is to say without time steps becoming too short.

We compare the stellar models evolved using each solution.
Even with sets of models with minimum numerical enhancement from each solution, the remnant mass of the stars can vary by up to 14\Msun{}, while the maximum radial expansion achieved by stars can vary by up to 2000\Rsun{} between the sets. 
These differences are important for comparing stellar models with observations and the possible feedback on the evolution of massive stars. For example, the abovementioned differences can have huge implications for studies involving binary interactions of stars and stellar remnants.

We also find that the differences in the various evolutionary properties predicted by the set of models computed with \revision{enhanced mixing length} and those using MLT++ are quite small. However, these models show a difference of 1000\,K in the minimum effective temperatures achieved by the star as a red supergiant (cf. Section~\ref{sec:mlt++}). The effective temperature is critical for determining the spectral properties and surface abundances of the stars \citep[][]{Davies2013}. Thus, this discrepancy in the effective temperature can be significant for studies involving the abundance properties of massive stars, such as galactic chemical composition studies.

In commonly used 1D stellar evolution models, a combination of these solutions is used and not necessarily limited to their minimum value. Thus, the differences in the evolutionary properties of massive stars can he higher than what we get here, as shown in Paper I.

Here we have considered an intermediate value of metallicity ($Z=$ \timestento{1.42}{-3}), as this is the metallicity where progenitors of current gravitational wave observations are more likely to form. Due to the dependence of opacity on chemical composition and the diminishing effect of opacity peaks at lower metallicities \citep{Sanyal:2017}, numerical issues related to the proximity to the Eddington limit become less prominent at lower metallicities. 
For example, using the same set of input parameters as the standard set (Section~\ref{sec:standard_model}), we find that stellar models fail to evolve for initial masses beyond 20\Msun{} at higher (solar) metallicity ($Z=$ \timestento{1.42}{-2}), but they evolve smoothly for initial masses up to 100\Msun{} at lower metallicity ($Z=$ \timestento{1.42}{-4}).

Multidimensional stellar models suggest that convection can be more turbulent and nonlocalized than assumed by MLT \citep{Kupka2009}. 
Three-dimensional modeling of the density inversion regions in massive stars by \citet{Jiang2015,Jiang2018} shows a complex interplay of convective and radiative transport dependent on the ratio of the photon diffusion time to the dynamical time and a smaller convective flux compared to 1D codes. There are ongoing efforts to improve the treatment of convective mixing in 1D codes using results from 3D simulations \citep[][]{Mosumgaard2018,Arnett2019,Schultz2020}.

As observations of massive stars become more accessible, and as our ability to compute 3D models of massive stars improves, the problem of density inversions might be resolved in the future. 
Meanwhile, it is important to be aware of and acknowledge the impact different pragmatic solutions can have on the evolutionary model sequences of massive stars. 

\begin{acknowledgements}
We thank Ross Church, Jan Eldridge, Debashis Sanyal, Matteo Cantiello, and Mathieu Renzo for useful comments and discussions. 
We also thank Pablo Marchant for the \href{https://github.com/orlox/mkipp}{script} to make Kippenhahn diagrams.
PA, JH and SS acknowledge the support by the Australian Research Council Centre of Excellence for Gravitational Wave Discovery (OzGrav), through project number CE170100004. 
SS is a recipient of an ARC Discovery Early Career Research Award (DE220100241). DSz has been supported by the Alexander von Humboldt Foundation.
This work made use of the OzSTAR high performance computer at Swinburne University of Technology. 
OzSTAR is funded by Swinburne University of Technology and the National Collaborative Research Infrastructure Strategy (NCRIS). This research was funded in part by the National Science
Center (NCN), Poland under grant number OPUS 2021/41/B/ST9/00757. For
the purpose of Open Access, the author has applied a CC-BY public copyright
license to any Author Accepted Manuscript (AAM) version arising from this
submission.
\end{acknowledgements}

%
  \bibliographystyle{aa} 
  \bibliography{references} 

\begin{thebibliography}{115}
\expandafter\ifx\csname natexlab\endcsname\relax\def\natexlab#1{#1}\fi

\bibitem[{{Abbott} {et~al.}(2016){Abbott}, {Abbott}, {Abbott}, {Abernathy},
  {Acernese}, {Ackley}, {Adams}, {Adams}, {Addesso}, {Adhikari}, {Adya}, \&
  et~al.}]{Abbott:2016}
{Abbott}, B.~P., {Abbott}, R., {Abbott}, T.~D., {et~al.} 2016, \prl, 116,
  061102

\bibitem[{{Abbott} {et~al.}(2017){Abbott}, {Abbott}, {Abbott}, {Acernese},
  {Ackley}, {Adams}, {Adams}, {Addesso}, {Adhikari}, {Adya}, \&
  et~al}]{Abbott:2017NSmerger}
{Abbott}, B.~P., {Abbott}, R., {Abbott}, T.~D., {et~al.} 2017, Physical Review
  Letters, 119, 161101

\bibitem[{{Agrawal} {et~al.}(2020){Agrawal}, {Hurley}, {Stevenson},
  {Sz{\'e}csi}, \& {Flynn}}]{Agrawal:2020}
{Agrawal}, P., {Hurley}, J., {Stevenson}, S., {Sz{\'e}csi}, D., \& {Flynn}, C.
  2020, \mnras, 497, 4549

\bibitem[{{Agrawal} {et~al.}(2022){Agrawal}, {Sz{\'e}csi}, {Stevenson},
  {Eldridge}, \& {Hurley}}]{Agrawal:2022}
{Agrawal}, P., {Sz{\'e}csi}, D., {Stevenson}, S., {Eldridge}, J.~J., \&
  {Hurley}, J. 2022, \mnras, 512, 5717

\bibitem[{{Alastuey} \& {Jancovici}(1978)}]{Alastuey1978}
{Alastuey}, A. \& {Jancovici}, B. 1978, \apj, 226, 1034

\bibitem[{{Alongi} {et~al.}(1993){Alongi}, {Bertelli}, {Bressan}, {Chiosi},
  {Fagotto}, {Greggio}, \& {Nasi}}]{Alongi1993}
{Alongi}, M., {Bertelli}, G., {Bressan}, A., {et~al.} 1993, \aaps, 97, 851

\bibitem[{{Arnett} {et~al.}(2019){Arnett}, {Meakin}, {Hirschi}, {Cristini},
  {Georgy}, {Campbell}, {Scott}, {Kaiser}, {Viallet}, \&
  {Moc{\'a}k}}]{Arnett2019}
{Arnett}, W.~D., {Meakin}, C., {Hirschi}, R., {et~al.} 2019, \apj, 882, 18

\bibitem[{{Asplund} {et~al.}(2009){Asplund}, {Grevesse}, {Sauval}, \&
  {Scott}}]{Asplund2009}
{Asplund}, M., {Grevesse}, N., {Sauval}, A.~J., \& {Scott}, P. 2009, \araa, 47,
  481

\bibitem[{{Beasor} {et~al.}(2020){Beasor}, {Davies}, {Smith}, {van Loon},
  {Gehrz}, \& {Figer}}]{Beasor2020}
{Beasor}, E.~R., {Davies}, B., {Smith}, N., {et~al.} 2020, \mnras, 492, 5994

\bibitem[{{Belczynski} {et~al.}(2010){Belczynski}, {Bulik}, {Fryer}, {Ruiter},
  {Valsecchi}, {Vink}, \& {Hurley}}]{Belczynski:2010}
{Belczynski}, K., {Bulik}, T., {Fryer}, C.~L., {et~al.} 2010, \apj, 714, 1217

\bibitem[{{Belczynski} {et~al.}(2008){Belczynski}, {Kalogera}, {Rasio}, {Taam},
  {Zezas}, {Bulik}, {Maccarone}, \& {Ivanova}}]{Belczynski:2008}
{Belczynski}, K., {Kalogera}, V., {Rasio}, F.~A., {et~al.} 2008, \apjs, 174,
  223

\bibitem[{{Bestenlehner} {et~al.}(2014){Bestenlehner}, {Gr{\"a}fener}, {Vink},
  {Najarro}, {de Koter}, {Sana}, {Evans}, {Crowther}, {H{\'e}nault-Brunet},
  {Herrero}, {Langer}, {Schneider}, {Sim{\'o}n-D{\'\i}az}, {Taylor}, \&
  {Walborn}}]{Bestenlehner2014}
{Bestenlehner}, J.~M., {Gr{\"a}fener}, G., {Vink}, J.~S., {et~al.} 2014, \aap,
  570, A38

\bibitem[{{Bj{\"o}rklund} {et~al.}(2021){Bj{\"o}rklund}, {Sundqvist}, {Puls},
  \& {Najarro}}]{Bjorklund:2021}
{Bj{\"o}rklund}, R., {Sundqvist}, J.~O., {Puls}, J., \& {Najarro}, F. 2021,
  \aap, 648, A36

\bibitem[{{B{\"o}hm-Vitense}(1958)}]{BohmVitense1958}
{B{\"o}hm-Vitense}, E. 1958, \zap, 46, 108

\bibitem[{{Bowman}(2020)}]{Bowman2020}
{Bowman}, D.~M. 2020, Frontiers in Astronomy and Space Sciences, 7, 70

\bibitem[{{Brott} {et~al.}(2011){Brott}, {de Mink}, {Cantiello}, {Langer}, {de
  Koter}, {Evans}, {Hunter}, {Trundle}, \& {Vink}}]{Brott:2011}
{Brott}, I., {de Mink}, S.~E., {Cantiello}, M., {et~al.} 2011, \aap, 530, A115

\bibitem[{{Cantiello} {et~al.}(2009){Cantiello}, {Langer}, {Brott}, {de Koter},
  {Shore}, {Vink}, {Voegler}, {Lennon}, \& {Yoon}}]{Cantiello2009}
{Cantiello}, M., {Langer}, N., {Brott}, I., {et~al.} 2009, \aap, 499, 279

\bibitem[{{Castro} {et~al.}(2018){Castro}, {Oey}, {Fossati}, \&
  {Langer}}]{Castro:2018}
{Castro}, N., {Oey}, M.~S., {Fossati}, L., \& {Langer}, N. 2018, \apj, 868, 57

\bibitem[{{Chen} {et~al.}(2015){Chen}, {Bressan}, {Girardi}, {Marigo}, {Kong},
  \& {Lanza}}]{Cheng:2015}
{Chen}, Y., {Bressan}, A., {Girardi}, L., {et~al.} 2015, \mnras, 452, 1068

\bibitem[{{Choi} {et~al.}(2016){Choi}, {Dotter}, {Conroy}, {Cantiello},
  {Paxton}, \& {Johnson}}]{Choi:2016MIST}
{Choi}, J., {Dotter}, A., {Conroy}, C., {et~al.} 2016, \apj, 823, 102

\bibitem[{{Cyburt} {et~al.}(2010){Cyburt}, {Amthor}, {Ferguson}, {Meisel},
  {Smith}, {Warren}, {Heger}, {Hoffman}, {Rauscher}, {Sakharuk}, {Schatz},
  {Thielemann}, \& {Wiescher}}]{Cyburt2010}
{Cyburt}, R.~H., {Amthor}, A.~M., {Ferguson}, R., {et~al.} 2010, \apjs, 189,
  240

\bibitem[{{Davies} {et~al.}(2018){Davies}, {Crowther}, \&
  {Beasor}}]{Davies:2018}
{Davies}, B., {Crowther}, P.~A., \& {Beasor}, E.~R. 2018, \mnras, 478, 3138

\bibitem[{{Davies} {et~al.}(2013){Davies}, {Kudritzki}, {Plez}, {Trager},
  {Lan{\c{c}}on}, {Gazak}, {Bergemann}, {Evans}, \& {Chiavassa}}]{Davies2013}
{Davies}, B., {Kudritzki}, R.-P., {Plez}, B., {et~al.} 2013, \apj, 767, 3

\bibitem[{{de Jager} {et~al.}(1988){de Jager}, {Nieuwenhuijzen}, \& {van der
  Hucht}}]{deJagerandNieuwenhuijzen:1988}
{de Jager}, C., {Nieuwenhuijzen}, H., \& {van der Hucht}, K.~A. 1988, \aaps,
  72, 259

\bibitem[{{Eddington}(1926)}]{Eddington1926}
{Eddington}, A.~S. 1926, {The Internal Constitution of the Stars}, Cambridge
  Science Classics (Cambridge University Press)

\bibitem[{{Ekstr{\"o}m} {et~al.}(2012){Ekstr{\"o}m}, {Georgy}, {Eggenberger},
  {Meynet}, {Mowlavi}, {Wyttenbach}, {Granada}, {Decressin}, {Hirschi},
  {Frischknecht}, {Charbonnel}, \& {Maeder}}]{Ekstroem:2012}
{Ekstr{\"o}m}, S., {Georgy}, C., {Eggenberger}, P., {et~al.} 2012, \aap, 537,
  A146

\bibitem[{{Ekstr{\"o}m} {et~al.}(2011){Ekstr{\"o}m}, {Georgy}, {Meynet},
  {Maeder}, \& {Granada}}]{Ekstroem:2011}
{Ekstr{\"o}m}, S., {Georgy}, C., {Meynet}, G., {Maeder}, A., \& {Granada}, A.
  2011, in IAU Symposium, Vol. 272, IAU Symposium, ed. C.~{Neiner}, G.~{Wade},
  G.~{Meynet}, \& G.~{Peters}, 62--72

\bibitem[{{Ekstr{\"o}m} {et~al.}(2020){Ekstr{\"o}m}, {Meynet}, {Georgy},
  {Hirschi}, {Maeder}, {Groh}, {Eggenberger}, \& {Buldgen}}]{Ekstroem2020}
{Ekstr{\"o}m}, S., {Meynet}, G., {Georgy}, C., {et~al.} 2020, in Stars and
  their Variability Observed from Space, ed. C.~{Neiner}, W.~W. {Weiss},
  D.~{Baade}, R.~E. {Griffin}, C.~C. {Lovekin}, \& A.~F.~J. {Moffat}, 223--228

\bibitem[{{Eldridge} \& {Tout}(2004)}]{Eldridge2004}
{Eldridge}, J.~J. \& {Tout}, C.~A. 2004, \mnras, 353, 87

\bibitem[{{{\'E}rgma}(1971)}]{Ergma1971}
{{\'E}rgma}, {\'E}. 1971, \sovast, 15, 51

\bibitem[{Ertl {et~al.}(2016)Ertl, Janka, Woosley, Sukhbold, \&
  Ugliano}]{Ertl:2015}
Ertl, T., Janka, H.~T., Woosley, S.~E., Sukhbold, T., \& Ugliano, M. 2016,
  Astrophys. J., 818, 124

\bibitem[{{Evans} {et~al.}(2011){Evans}, {Taylor}, {H{\'e}nault-Brunet},
  {Sana}, {de Koter}, {Sim{\'o}n-D{\'\i}az}, {Carraro}, {Bagnoli}, {Bastian},
  {Bestenlehner}, {Bonanos}, {Bressert}, {Brott}, {Campbell}, {Cantiello},
  {Clark}, {Costa}, {Crowther}, {de Mink}, {Doran}, {Dufton}, {Dunstall},
  {Friedrich}, {Garcia}, {Gieles}, {Gr{\"a}fener}, {Herrero}, {Howarth},
  {Izzard}, {Langer}, {Lennon}, {Ma{\'\i}z Apell{\'a}niz}, {Markova},
  {Najarro}, {Puls}, {Ramirez}, {Sab{\'\i}n-Sanjuli{\'a}n}, {Smartt}, {Stroud},
  {van Loon}, {Vink}, \& {Walborn}}]{Evans:2011}
{Evans}, C.~J., {Taylor}, W.~D., {H{\'e}nault-Brunet}, V., {et~al.} 2011, \aap,
  530, A108

\bibitem[{{Farmer} {et~al.}(2016){Farmer}, {Fields}, {Petermann}, {Dessart},
  {Cantiello}, {Paxton}, \& {Timmes}}]{Farmer:2016}
{Farmer}, R., {Fields}, C.~E., {Petermann}, I., {et~al.} 2016, \apjs, 227, 22

\bibitem[{{Fields} {et~al.}(2018){Fields}, {Timmes}, {Farmer}, {Petermann},
  {Wolf}, \& {Couch}}]{Fields:2018}
{Fields}, C.~E., {Timmes}, F.~X., {Farmer}, R., {et~al.} 2018, \apjs, 234, 19

\bibitem[{{Fryer} {et~al.}(2012){Fryer}, {Belczynski}, {Wiktorowicz},
  {Dominik}, {Kalogera}, \& {Holz}}]{Fryer:2012}
{Fryer}, C.~L., {Belczynski}, K., {Wiktorowicz}, G., {et~al.} 2012, \apj, 749,
  91

\bibitem[{{Gilkis} {et~al.}(2021){Gilkis}, {Shenar}, {Ramachandran}, {Jermyn},
  {Mahy}, {Oskinova}, {Arcavi}, \& {Sana}}]{Gilkis2021}
{Gilkis}, A., {Shenar}, T., {Ramachandran}, V., {et~al.} 2021, \mnras, 503,
  1884

\bibitem[{{Glebbeek} {et~al.}(2009){Glebbeek}, {Gaburov}, {de Mink}, {Pols}, \&
  {Portegies Zwart}}]{Glebbeek:2009}
{Glebbeek}, E., {Gaburov}, E., {de Mink}, S.~E., {Pols}, O.~R., \& {Portegies
  Zwart}, S.~F. 2009, \aap, 497, 255

\bibitem[{{Graboske} {et~al.}(1973){Graboske}, {Dewitt}, {Grossman}, \&
  {Cooper}}]{Graboske1973}
{Graboske}, H.~C., {Dewitt}, H.~E., {Grossman}, A.~S., \& {Cooper}, M.~S. 1973,
  \apj, 181, 457

\bibitem[{{Gr{\"a}fener}(2021)}]{Grafener:2021}
{Gr{\"a}fener}, G. 2021, \aap, 647, A13

\bibitem[{{Gr{\"a}fener} \& {Hamann}(2005)}]{Grafener:2005}
{Gr{\"a}fener}, G. \& {Hamann}, W.~R. 2005, \aap, 432, 633

\bibitem[{{Gr{\"a}fener} \& {Hamann}(2008)}]{Grafener:2008}
{Gr{\"a}fener}, G. \& {Hamann}, W.~R. 2008, \aap, 482, 945

\bibitem[{{Gr{\"a}fener} {et~al.}(2012){Gr{\"a}fener}, {Owocki}, \&
  {Vink}}]{Grafener:2012}
{Gr{\"a}fener}, G., {Owocki}, S.~P., \& {Vink}, J.~S. 2012, \aap, 538, A40

\bibitem[{{Groh} {et~al.}(2020){Groh}, {Farrell}, {Meynet}, {Smith}, {Murphy},
  {Allan}, {Georgy}, \& {Ekstroem}}]{Groh:2020}
{Groh}, J.~H., {Farrell}, E.~J., {Meynet}, G., {et~al.} 2020, \apj, 900, 98

\bibitem[{{Hawcroft} {et~al.}(2021){Hawcroft}, {Sana}, {Mahy}, {Sundqvist},
  {Abdul-Masih}, {Bouret}, {Brands}, {de Koter}, {Driessen}, \&
  {Puls}}]{Hawcroft:2021A&A}
{Hawcroft}, C., {Sana}, H., {Mahy}, L., {et~al.} 2021, \aap, 655, A67

\bibitem[{{Heger} {et~al.}(2000){Heger}, {Langer}, \& {Woosley}}]{Heger:2000a}
{Heger}, A., {Langer}, N., \& {Woosley}, S.~E. 2000, \apj, 528, 368

\bibitem[{{Heger} {et~al.}(2002){Heger}, {Woosley}, {Rauscher}, {Hoffman}, \&
  {Boyes}}]{Heger:2002}
{Heger}, A., {Woosley}, S.~E., {Rauscher}, T., {Hoffman}, R.~D., \& {Boyes},
  M.~M. 2002, \nar, 46, 463

\bibitem[{{Henyey} {et~al.}(1965){Henyey}, {Vardya}, \&
  {Bodenheimer}}]{Henyey1965}
{Henyey}, L., {Vardya}, M.~S., \& {Bodenheimer}, P. 1965, \apj, 142, 841

\bibitem[{Higgins \& Vink(2020)}]{Higgins:2020njv}
Higgins, E.~R. \& Vink, J.~S. 2020, Astron. Astrophys., 635, A175

\bibitem[{{Humphreys} \& {Davidson}(1979)}]{Humphreys:1979}
{Humphreys}, R.~M. \& {Davidson}, K. 1979, \apj, 232, 409

\bibitem[{{Humphreys} \& {Davidson}(1994)}]{HumphreysDavidson1994}
{Humphreys}, R.~M. \& {Davidson}, K. 1994, \pasp, 106, 1025

\bibitem[{{Iglesias} \& {Rogers}(1993)}]{Iglesias1993}
{Iglesias}, C.~A. \& {Rogers}, F.~J. 1993, \apj, 412, 752

\bibitem[{{Iglesias} \& {Rogers}(1996)}]{Iglesias1996}
{Iglesias}, C.~A. \& {Rogers}, F.~J. 1996, \apj, 464, 943

\bibitem[{{Ishii} {et~al.}(1999){Ishii}, {Ueno}, \& {Kato}}]{Ishii:1999}
{Ishii}, M., {Ueno}, M., \& {Kato}, M. 1999, \pasj, 51, 417

\bibitem[{{Itoh} {et~al.}(1979){Itoh}, {Totsuji}, {Ichimaru}, \&
  {Dewitt}}]{Itoh1979}
{Itoh}, N., {Totsuji}, H., {Ichimaru}, S., \& {Dewitt}, H.~E. 1979, \apj, 234,
  1079

\bibitem[{{Jermyn} {et~al.}(2022){Jermyn}, {Bauer}, {Schwab}, {Farmer}, {Ball},
  {Bellinger}, {Dotter}, {Joyce}, {Marchant}, {Mombarg}, {Wolf}, {Wong},
  {Cinquegrana}, {Farrell}, {Smolec}, {Thoul}, {Cantiello}, {Herwig}, {Toloza},
  {Bildsten}, {Townsend}, \& {Timmes}}]{Jermyn2022}
{Jermyn}, A.~S., {Bauer}, E.~B., {Schwab}, J., {et~al.} 2022, arXiv e-prints,
  arXiv:2208.03651

\bibitem[{{Jiang} {et~al.}(2015){Jiang}, {Cantiello}, {Bildsten}, {Quataert},
  \& {Blaes}}]{Jiang2015}
{Jiang}, Y.-F., {Cantiello}, M., {Bildsten}, L., {Quataert}, E., \& {Blaes}, O.
  2015, \apj, 813, 74

\bibitem[{{Jiang} {et~al.}(2018){Jiang}, {Cantiello}, {Bildsten}, {Quataert},
  {Blaes}, \& {Stone}}]{Jiang2018}
{Jiang}, Y.-F., {Cantiello}, M., {Bildsten}, L., {et~al.} 2018, \nat, 561, 498

\bibitem[{{Joss} {et~al.}(1973){Joss}, {Salpeter}, \& {Ostriker}}]{Joss1973}
{Joss}, P.~C., {Salpeter}, E.~E., \& {Ostriker}, J.~P. 1973, \apj, 181, 429

\bibitem[{{Kaiser} {et~al.}(2020){Kaiser}, {Hirschi}, {Arnett}, {Georgy},
  {Scott}, \& {Cristini}}]{Kaiser:2020}
{Kaiser}, E.~A., {Hirschi}, R., {Arnett}, W.~D., {et~al.} 2020, \mnras, 496,
  1967

\bibitem[{{Kasen} {et~al.}(2017){Kasen}, {Metzger}, {Barnes}, {Quataert}, \&
  {Ramirez-Ruiz}}]{Kasen2017}
{Kasen}, D., {Metzger}, B., {Barnes}, J., {Quataert}, E., \& {Ramirez-Ruiz}, E.
  2017, \nat, 551, 80

\bibitem[{{Kippenhahn} {et~al.}(1980){Kippenhahn}, {Ruschenplatt}, \&
  {Thomas}}]{Kippenhahn1980}
{Kippenhahn}, R., {Ruschenplatt}, G., \& {Thomas}, H.~C. 1980, \aap, 91, 175

\bibitem[{{Kippenhahn} {et~al.}(2012){Kippenhahn}, {Weigert}, \&
  {Weiss}}]{Kippenhahn2012}
{Kippenhahn}, R., {Weigert}, A., \& {Weiss}, A. 2012, {Stellar Structure and
  Evolution} (Springer)

\bibitem[{{Klencki} {et~al.}(2018){Klencki}, {Moe}, {Gladysz}, {Chruslinska},
  {Holz}, \& {Belczynski}}]{Klencki:2018}
{Klencki}, J., {Moe}, M., {Gladysz}, W., {et~al.} 2018, \aap, 619, A77

\bibitem[{{K{\"o}hler} {et~al.}(2015){K{\"o}hler}, {Langer}, {de Koter}, {de
  Mink}, {Crowther}, {Evans}, {Gr{\"a}fener}, {Sana}, {Sanyal}, {Schneider}, \&
  {Vink}}]{Kohler:2015}
{K{\"o}hler}, K., {Langer}, N., {de Koter}, A., {et~al.} 2015, \aap, 573, A71

\bibitem[{{Kupka}(2009)}]{Kupka2009}
{Kupka}, F. 2009, {Turbulent Convection and Numerical Simulations in Solar and
  Stellar Astrophysics}, Vol. 756 (Springer Berlin Heidelberg), 49

\bibitem[{{Lamers} \& {Fitzpatrick}(1988)}]{Lamers1988}
{Lamers}, H. J.~G.~L.~M. \& {Fitzpatrick}, E.~L. 1988, \apj, 324, 279

\bibitem[{{Langer}(1997)}]{Langer1997}
{Langer}, N. 1997, in Astronomical Society of the Pacific Conference Series,
  Vol. 120, Luminous Blue Variables: Massive Stars in Transition, ed. A.~{Nota}
  \& H.~{Lamers}, 83

\bibitem[{{Langer}(2012)}]{Langer:2012}
{Langer}, N. 2012, \araa, 50, 107

\bibitem[{{Langer} {et~al.}(1985){Langer}, {El Eid}, \& {Fricke}}]{Langer1985}
{Langer}, N., {El Eid}, M.~F., \& {Fricke}, K.~J. 1985, \aap, 145, 179

\bibitem[{{Maeder}(1987)}]{Maeder1987}
{Maeder}, A. 1987, \aap, 173, 247

\bibitem[{{Maeder}(1992)}]{Maeder:1992}
{Maeder}, A. 1992, in Instabilities in Evolved Super- and Hypergiants, ed.
  C.~{de Jager} \& H.~{Nieuwenhuijzen}, 138

\bibitem[{{Maeder}(2009)}]{Maeder2009}
{Maeder}, A. 2009, {Physics, Formation and Evolution of Rotating Stars}
  (Springer Berlin Heidelberg)

\bibitem[{{Mihalas}(1969)}]{Mihalas:1969}
{Mihalas}, D. 1969, \apjl, 156, L155

\bibitem[{{Mosumgaard} {et~al.}(2018){Mosumgaard}, {Ball}, {Silva Aguirre},
  {Weiss}, \& {Christensen-Dalsgaard}}]{Mosumgaard2018}
{Mosumgaard}, J.~R., {Ball}, W.~H., {Silva Aguirre}, V., {Weiss}, A., \&
  {Christensen-Dalsgaard}, J. 2018, \mnras, 478, 5650

\bibitem[{{Nabi} {et~al.}(2019){Nabi}, {Shehzadi}, \& {Majid}}]{Nabi2019}
{Nabi}, J.-U., {Shehzadi}, R., \& {Majid}, M. 2019, \na, 71, 12

\bibitem[{{Nugis} \& {Lamers}(2000)}]{NugisandLamers:2000}
{Nugis}, T. \& {Lamers}, H.~J.~G.~L.~M. 2000, \aap, 360, 227

\bibitem[{{O'Connor} \& {Ott}(2011)}]{OConnor:2011}
{O'Connor}, E. \& {Ott}, C.~D. 2011, \apj, 730, 70

\bibitem[{{Owocki}(2015)}]{Owocki2015}
{Owocki}, S.~P. 2015, {Instabilities in the Envelopes and Winds of Very Massive
  Stars}, Vol. 412 (Springer International Publishing), 113

\bibitem[{{Owocki} {et~al.}(2004){Owocki}, {Gayley}, \& {Shaviv}}]{Owocki2004}
{Owocki}, S.~P., {Gayley}, K.~G., \& {Shaviv}, N.~J. 2004, \apj, 616, 525

\bibitem[{{Paxton} {et~al.}(2011){Paxton}, {Bildsten}, {Dotter}, {Herwig},
  {Lesaffre}, \& {Timmes}}]{Paxton2011}
{Paxton}, B., {Bildsten}, L., {Dotter}, A., {et~al.} 2011, \apjs, 192, 3

\bibitem[{{Paxton} {et~al.}(2013){Paxton}, {Cantiello}, {Arras}, {Bildsten},
  {Brown}, {Dotter}, {Mankovich}, {Montgomery}, {Stello}, {Timmes}, \&
  {Townsend}}]{Paxton2013}
{Paxton}, B., {Cantiello}, M., {Arras}, P., {et~al.} 2013, \apjs, 208, 4

\bibitem[{{Paxton} {et~al.}(2015){Paxton}, {Marchant}, {Schwab}, {Bauer},
  {Bildsten}, {Cantiello}, {Dessart}, {Farmer}, {Hu}, {Langer}, {Townsend},
  {Townsley}, \& {Timmes}}]{Paxton2015}
{Paxton}, B., {Marchant}, P., {Schwab}, J., {et~al.} 2015, \apjs, 220, 15

\bibitem[{{Paxton} {et~al.}(2018){Paxton}, {Schwab}, {Bauer}, {Bildsten},
  {Blinnikov}, {Duffell}, {Farmer}, {Goldberg}, {Marchant}, {Sorokina},
  {Thoul}, {Townsend}, \& {Timmes}}]{Paxton2018}
{Paxton}, B., {Schwab}, J., {Bauer}, E.~B., {et~al.} 2018, \apjs, 234, 34

\bibitem[{{Paxton} {et~al.}(2019){Paxton}, {Smolec}, {Schwab}, {Gautschy},
  {Bildsten}, {Cantiello}, {Dotter}, {Farmer}, {Goldberg}, {Jermyn}, {Kanbur},
  {Marchant}, {Thoul}, {Townsend}, {Wolf}, {Zhang}, \& {Timmes}}]{Paxton2019}
{Paxton}, B., {Smolec}, R., {Schwab}, J., {et~al.} 2019, \apjs, 243, 10

\bibitem[{{Pedersen} {et~al.}(2019){Pedersen}, {Chowdhury}, {Johnston},
  {Bowman}, {Aerts}, {Handler}, {De Cat}, {Neiner}, {David-Uraz}, {Buzasi},
  {Tkachenko}, {Sim{\'o}n-D{\'\i}az}, {Moravveji}, {Sikora}, {Mirouh},
  {Lovekin}, {Cantiello}, {Daszy{\'n}ska-Daszkiewicz}, {Pigulski},
  {Vanderspek}, \& {Ricker}}]{Pedersen2019}
{Pedersen}, M.~G., {Chowdhury}, S., {Johnston}, C., {et~al.} 2019, \apjl, 872,
  L9

\bibitem[{{Petrovic} {et~al.}(2006){Petrovic}, {Pols}, \&
  {Langer}}]{Petrovic:2006}
{Petrovic}, J., {Pols}, O., \& {Langer}, N. 2006, \aap, 450, 219

\bibitem[{{Puls} {et~al.}(2008){Puls}, {Markova}, \& {Scuderi}}]{Puls2008a}
{Puls}, J., {Markova}, N., \& {Scuderi}, S. 2008, in Astronomical Society of
  the Pacific Conference Series, Vol. 388, Mass Loss from Stars and the
  Evolution of Stellar Clusters, ed. A.~{de Koter}, L.~J. {Smith}, \&
  L.~B.~F.~M. {Waters}, 101

\bibitem[{{Renzini}(1987)}]{Renzini1987}
{Renzini}, A. 1987, \aap, 188, 49

\bibitem[{{Renzo} {et~al.}(2017){Renzo}, {Ott}, {Shore}, \& {de
  Mink}}]{Renzo:2017}
{Renzo}, M., {Ott}, C.~D., {Shore}, S.~N., \& {de Mink}, S.~E. 2017, \aap, 603,
  A118

\bibitem[{{Sabhahit} {et~al.}(2021){Sabhahit}, {Vink}, {Higgins}, \&
  {Sander}}]{Sabhahit:2021}
{Sabhahit}, G.~N., {Vink}, J.~S., {Higgins}, E.~R., \& {Sander}, A. A.~C. 2021,
  \mnras, 506, 4473

\bibitem[{{Sabhahit} {et~al.}(2022){Sabhahit}, {Vink}, {Higgins}, \&
  {Sander}}]{Sabhahit:2022}
{Sabhahit}, G.~N., {Vink}, J.~S., {Higgins}, E.~R., \& {Sander}, A. A.~C. 2022,
  \mnras, 514, 3736

\bibitem[{{Saio} {et~al.}(1998){Saio}, {Baker}, \& {Gautschy}}]{Saio1998}
{Saio}, H., {Baker}, N.~H., \& {Gautschy}, A. 1998, \mnras, 294, 622

\bibitem[{{Saio} {et~al.}(2013){Saio}, {Georgy}, \& {Meynet}}]{Saio2013}
{Saio}, H., {Georgy}, C., \& {Meynet}, G. 2013, in Astronomical Society of the
  Pacific Conference Series, Vol. 479, Progress in Physics of the Sun and
  Stars: A New Era in Helio- and Asteroseismology, ed. H.~{Shibahashi} \& A.~E.
  {Lynas-Gray}, 47

\bibitem[{{Sanyal} {et~al.}(2015){Sanyal}, {Grassitelli}, {Langer}, \&
  {Bestenlehner}}]{Sanyal:2015}
{Sanyal}, D., {Grassitelli}, L., {Langer}, N., \& {Bestenlehner}, J.~M. 2015,
  \aap, 580, A20

\bibitem[{{Sanyal} {et~al.}(2017){Sanyal}, {Langer}, {Sz{\'e}csi}, {-C Yoon},
  \& {Grassitelli}}]{Sanyal:2017}
{Sanyal}, D., {Langer}, N., {Sz{\'e}csi}, D., {-C Yoon}, S., \& {Grassitelli},
  L. 2017, \aap, 597, A71

\bibitem[{{Schootemeijer} {et~al.}(2019){Schootemeijer}, {Langer}, {Grin}, \&
  {Wang}}]{Schootemeijer:2019}
{Schootemeijer}, A., {Langer}, N., {Grin}, N.~J., \& {Wang}, C. 2019, \aap,
  625, A132

\bibitem[{{Schultz} {et~al.}(2020){Schultz}, {Bildsten}, \&
  {Jiang}}]{Schultz2020}
{Schultz}, W.~C., {Bildsten}, L., \& {Jiang}, Y.-F. 2020, \apj, 902, 67

\bibitem[{{Schwab}(2019)}]{Schwab2019}
{Schwab}, J. 2019, \apj, 885, 27

\bibitem[{{Sim{\'o}n-D{\'\i}az} {et~al.}(2017){Sim{\'o}n-D{\'\i}az}, {Godart},
  {Castro}, {Herrero}, {Aerts}, {Puls}, {Telting}, \&
  {Grassitelli}}]{Simon-Diaz2017}
{Sim{\'o}n-D{\'\i}az}, S., {Godart}, M., {Castro}, N., {et~al.} 2017, \aap,
  597, A22

\bibitem[{{Sim{\'o}n-D{\'\i}az} \& {Herrero}(2014)}]{simon-diaz2014}
{Sim{\'o}n-D{\'\i}az}, S. \& {Herrero}, A. 2014, \aap, 562, A135

\bibitem[{{Sim{\'o}n-D{\'\i}az} {et~al.}(2015){Sim{\'o}n-D{\'\i}az},
  {Negueruela}, {Ma{\'\i}z Apell{\'a}niz}, {Castro}, {Herrero}, {Garcia},
  {P{\'e}rez-Prieto}, {Caon}, {Alacid}, {Camacho}, {Dorda}, {Godart},
  {Gonz{\'a}lez-Fern{\'a}ndez}, {Holgado}, \&
  {R{\"u}bke}}]{Simon-Diaz2015IACOB}
{Sim{\'o}n-D{\'\i}az}, S., {Negueruela}, I., {Ma{\'\i}z Apell{\'a}niz}, J.,
  {et~al.} 2015, in Highlights of Spanish Astrophysics VIII, 576--581

\bibitem[{{Smith}(2014)}]{Smith:2014}
{Smith}, N. 2014, \araa, 52, 487

\bibitem[{{Smith} \& {Owocki}(2006)}]{Smith:2006ApJL}
{Smith}, N. \& {Owocki}, S.~P. 2006, \apjl, 645, L45

\bibitem[{{Smith} \& {Tombleson}(2015)}]{Smith:2015}
{Smith}, N. \& {Tombleson}, R. 2015, \mnras, 447, 598

\bibitem[{{Smith} {et~al.}(2004){Smith}, {Vink}, \& {de Koter}}]{Smith:2004}
{Smith}, N., {Vink}, J.~S., \& {de Koter}, A. 2004, \apj, 615, 475

\bibitem[{{Stevenson} {et~al.}(2017){Stevenson}, {Vigna-G{\'o}mez}, {Mandel},
  {Barrett}, {Neijssel}, {Perkins}, \& {de Mink}}]{Stevenson:2017}
{Stevenson}, S., {Vigna-G{\'o}mez}, A., {Mandel}, I., {et~al.} 2017, Nature
  Communications, 8, 14906

\bibitem[{{Stothers} \& {Chin}(1979)}]{Stothers:1979}
{Stothers}, R. \& {Chin}, C.~W. 1979, \apj, 233, 267

\bibitem[{{Stothers} \& {Chin}(1993)}]{Stothers:1993}
{Stothers}, R.~B. \& {Chin}, C.-W. 1993, \apjl, 408, L85

\bibitem[{{Vink}(2011)}]{Vink:2011}
{Vink}, J.~S. 2011, \apss, 336, 163

\bibitem[{{Vink} \& {de Koter}(2005)}]{Vink:2005}
{Vink}, J.~S. \& {de Koter}, A. 2005, \aap, 442, 587

\bibitem[{{Vink} {et~al.}(2000){Vink}, {de Koter}, \& {Lamers}}]{Vink:2000}
{Vink}, J.~S., {de Koter}, A., \& {Lamers}, H.~J.~G.~L.~M. 2000, \aap, 362, 295

\bibitem[{{Vink} {et~al.}(2001){Vink}, {de Koter}, \& {Lamers}}]{Vink:2001}
{Vink}, J.~S., {de Koter}, A., \& {Lamers}, H.~J.~G.~L.~M. 2001, \aap, 369, 574

\bibitem[{{Vink} {et~al.}(2021){Vink}, {Higgins}, {Sander}, \&
  {Sabhahit}}]{Vink2021b}
{Vink}, J.~S., {Higgins}, E.~R., {Sander}, A. A.~C., \& {Sabhahit}, G.~N. 2021,
  \mnras, 504, 146

\bibitem[{{Vink} \& {Sander}(2021)}]{Vink2021a}
{Vink}, J.~S. \& {Sander}, A. A.~C. 2021, \mnras, 504, 2051

\bibitem[{{Wade} {et~al.}(2014){Wade}, {Grunhut}, {Alecian}, {Neiner},
  {Auri{\`e}re}, {Bohlender}, {David-Uraz}, {Folsom}, {Henrichs}, {Kochukhov},
  {Mathis}, {Owocki}, {Petit}, \& {Petit}}]{Wade2014MiMes}
{Wade}, G.~A., {Grunhut}, J., {Alecian}, E., {et~al.} 2014, in Magnetic Fields
  throughout Stellar Evolution, Vol. 302, Magnetic Fields throughout Stellar
  Evolution, ed. P.~{Petit}, M.~{Jardine}, \& H.~C. {Spruit}, 265--269

\end{thebibliography}
%


\begin{appendix} 

\section{Details of MLT++}

\label{subsec:details_MLT++}
In MLT++, \textsc{MESA} makes convection efficient by 
reducing superadiabaticity in the stellar envelope. 
In the regions where superadiabaticity is greater than threshold \verb|gradT_excess_f1|, it decreases ${\rm \nabla_T}$ to make it closer to ${\rm \nabla_{ad}}$. 
The fraction of decrease is determined by the parameter \verb|gradT_excess_alpha| or $\alpha$ and is calculated based on the value of \citep[cf. equation 38 of][]{Paxton2013}:
\begin{equation}
\lambda_{\max} \equiv \max \left(\frac{L_{\mathrm{rad}}}{L_{\mathrm{Edd}}}\right) \quad \text { and } \quad \beta_{\min} \equiv \min \left(\frac{P_{\mathrm{gas}}}{P}\right) \, .\end{equation}

For each stellar model, MESA computes $\alpha$ by comparing $\lambda_{\max}$ with the thresholds $\lambda_{1}$ and $\lambda_{2}$ and $\beta_{\min}$ with $\beta_{1}$ and $\beta_{2}$ using the following conditions. 

If $\lambda_{\max} \geq \lambda_{1}$, then
\begin{align}
\alpha=
\begin{cases} 
      1 & \beta_{\rm min}\ \leq \beta_{1} \\
      \cfrac{\beta_{1}+d \beta-\beta_{\rm min}}{d \beta} & \beta_{1}< \beta_{\rm min} < \beta_{1}+d \beta\\
      0 & \text{otherwise} \\
   \end{cases}.
   \label{eq:mlt1}
\end{align}

If $\lambda_{\max} \geq \lambda_{2}$, then
\begin{align}
\alpha=
\begin{cases} 
      1 & \beta_{\rm min}\ \leq \beta_{\rm limit} \\
      \cfrac{\beta_{\rm limit}+d \beta-\beta_{\rm min}}{d \beta} & \beta_{\rm min} < \beta_{\rm limit}+d \beta\\
      0 & \beta_{\rm min} \geq \beta_{\rm limit}+d \beta \\
   \end{cases}.
\end{align}

If $\lambda_{\max} > \lambda_{2}-d \lambda$, then
\begin{align}
\alpha=
\begin{cases} 
      1 & \beta_{\rm min}\ \leq \beta_{2} \\
      \cfrac{\lambda_{\rm max}+d \lambda-\lambda_{2}}{d \lambda} & \beta_{2}< \beta_{\rm min} < \beta_{2}+d \beta\\
      0 & \text{otherwise} \\
   \end{cases}.
   \label{eq:mlt3}
\end{align}

The net fraction of decrease is determined by a combination of alpha and user defined \verb|gradT_excess_f2| or $f_2$ using the following equation,

\begin{equation}
    {\rm \alpha_{net}} = f_2+ (1 - f_2)(1 - \alpha).
\end{equation}

The excess fraction is then subtracted from ${\rm \nabla_T}$, to give reduced ${\rm \nabla_{T,new}}$ as, 

\begin{equation}
    {\rm \nabla_{T,new}  = \alpha_{net} \times \nabla_T + (1-\alpha_{net}) \times \nabla_{ad}}.
    \label{eq:mlt_final}
\end{equation}

The default values for the different thresholds in the Equations\,\ref{eq:mlt1}--\ref{eq:mlt3} are: $\lambda_{1}=1.0$ and  $\lambda_{2}=0.5$, $\beta_{1}=0.35$ and $\beta_{2} =0.25$, $d \lambda =0.1$ and $d \beta=0.1$. \verb|gradT_excess_f1| defaults to \tento{-4} while the default value of \verb|gradT_excess_f2| or $f_2$  is \tento{-3}. These values can be redefined by the user.

Equations\,\ref{eq:mlt1}--\ref{eq:mlt3} yield a value of $\alpha$ between 0 and 1. 
The maximum fraction of ${\rm \nabla_T}$ used in calculating ${\rm \nabla_{T,new}}$ is limited to $f_2$ (when $\alpha=1$), with smaller $f_2$ implying larger contribution of ${\rm \nabla_{ad}}$ in the equation\,\ref{eq:mlt_final} and therefore larger reduction in superadiabaticity.

\begin{figure}
    \centering
    \includegraphics[width=\columnwidth]{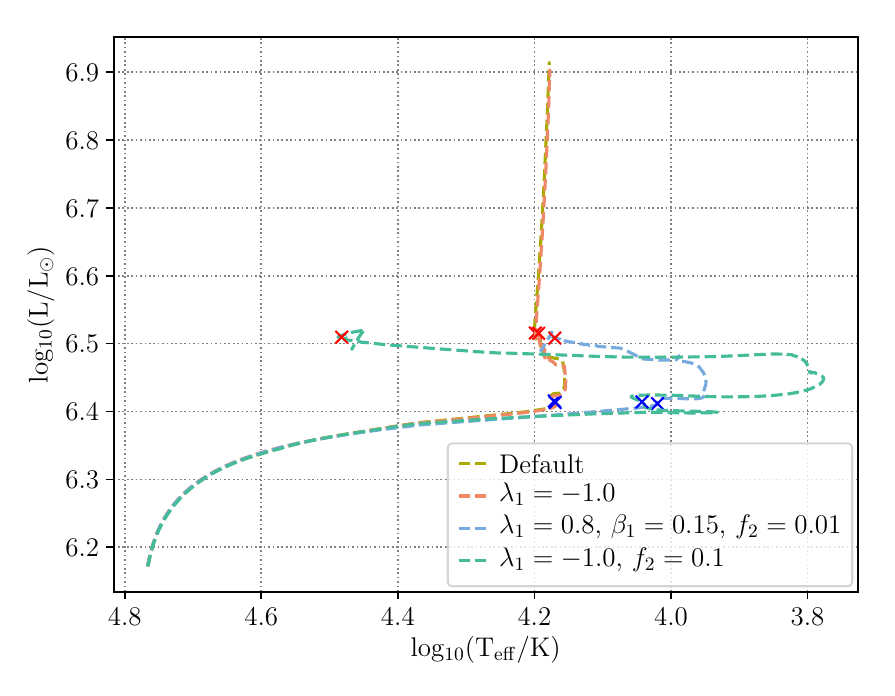}
    \caption{Evolutionary tracks of a 110\Msun{} star evolved with four different MLT++ parameter combinations. Apart from the changes indicated in the figure legend, default values for all other MLT++ parameters have been used. A negative value of $\lambda_{1}$ implies that the maximum reduction in superadiabaticity is applied throughout the track. The tracks corresponding to the default values of MLT++ and to $\lambda_{1}=-1.0$ are indistinguishable here and predict unrealistic luminosities.}
    \label{fig:mlt_test}
    
\end{figure}

We tested our models with for different values of $f_2$, $\lambda_{1}$ and $\beta_{1}$. The HR diagram for 110\Msun{} stellar model, calculated with the four different combinations of parameters in MLT++ is shown in Figure~\ref{fig:mlt_test}. We find that using a small reduction in superadiabaticity with $f_2$=\tento{-1} and setting $\lambda_{1} = 0.6$ and $\beta_{1}= 0.05$ helps the stellar models reach completion timely and without any numerical inaccuracies or difficulty.

\end{appendix}

\end{document}